%% file: main.tex
\documentclass[11pt]{article}
\input{headers}
\EnableStatementPartReferences

\declaretheorem[sibling=theorem]{definition}
\DeclareMathOperator{\size}{size}
\DeclareMathOperator{\Sym}{Sym}

\title{Sketching Language Models for Context Attribution}
\author{
    \begin{tabular}{c@{\hspace{2cm}}c}
        {\bf Christopher Liaw}
            & {\bf Aranyak Mehta}\\
        Google Research
            & Google Research\\[1.5ex]
        {\bf Grigoris Velegkas}
            & {\bf Felix Zhou}\\
        Google Research
            & Yale University\thanks{Part of this work was completed during an internship at Google Research.}
    \end{tabular}
}
\date{}

\begin{document}
\pagestyle{empty}
\pagenumbering{roman}

\maketitle
\thispagestyle{empty}

\begin{abstract}\noindent
Context attribution asks how removing parts of a prompt changes
a language model's output.
We study leave-one-out (LOO) attribution for a prompt of $n$
tokens and a fixed measurement function.
Computing all LOO values directly requires $n$ model evaluations.
Under a stability assumption, our algorithm estimates all LOO values
to error proportional to $\eps$ with constant probability using
$O(\sqrt{\log(\nicefrac1\eps)}/\eps)$ evaluation calls.
To complement our upper bound, we prove a matching
$\Omega(\sqrt{\log(\nicefrac1\eps)}/\eps)$ lower bound on the number of
evaluation calls in an oracle model.
Empirically, we verify coefficient decay consistent with stability
and evaluate the performance of our sketching algorithm on a
small language-model retrieval task.

Technically, our result is based on a sketching algorithm for
arithmetic circuits whose Taylor coefficients satisfy the stability
condition recently introduced by \citet{gunn2026sketch}.
A direct adaptation of their forward differentiation sketches
would require $\widetilde O(\log(n)/\eps^2)$
evaluation calls.
Our key observation is that we can apply forward differentiation
to the backpropagation circuit, with only constant overhead over
a forward sketch of the same Taylor degree.
This reduces the variance of the sketches and is key to our
improved evaluation bound.
Our lower bound builds on the techniques of \citet*[SIAM Journal on Matrix Analysis and Applications]{amsel2024fixed}
for lower bounding the number of matrix--vector queries to estimate the diagonal elements of random matrices.
\end{abstract}

\clearpage
\tableofcontents

\clearpage
\pagestyle{plain}
\pagenumbering{arabic}

\section{Introduction}
\label{sec:introduction}
How can we reliably determine which pieces of context actually
contributed to a model's output?
Frontier AI models frequently process documents and external tool
outputs in context to improve output quality
(\eg{}, search results). 
Answering this question helps us assign credit to sources,
identify context that can be pruned, and even detect certain poisoning
attacks~\citep{cohenwang2024contextcite}.

Formally, the problem is to counterfactually estimate how removing parts of the context would have affected the model behavior~\citep{liu2025attribot}.
To be more precise, assume that we are given access to a deterministic measurement function $\phi$ whose evaluation involves running an expensive model and outputting some scalar metric.
For instance, if $(\tau_1, \dots, \tau_n)$ is a sequence of $n$ tokens
forming a prompt to a language model,
and $z=(z_1, \dots, z_m)$ are the generated response tokens,
then $\phi(\tau_1, \dots, \tau_n) = \phi_z(\tau_1, \dots, \tau_n)$ might be the log-probability that the model would have generated $z$ using the prompt $(\tau_1,\ldots,\tau_n)$.

In this work, we focus on \emph{leave-one-out (LOO)} context attribution,
a well-studied and important setting in practice~\citep{li2016understanding,cohenwang2024contextcite,liu2025attribot}.
That is, we quantify the importance of the $i$-th token by the value of $\phi$ without the $i$-th token\footnote{Our techniques extend to the setting where there is an arbitrary partition of the tokens and we wish to estimate the value of $\phi$ after deleting a set of tokens in the partition.}
\[
    \phi(\tau_1, \dots, \tau_{i-1}, \tau_{i+1}, \dots, \tau_n)\,.
\]
Note that there is a trivial algorithm that exactly computes all LOO values using $n$ different evaluations of $\phi$,
but is likely to be intractable for large $n$.
Our goal is thus to estimate all LOO values with fewer evaluation calls.

\begin{framed}
Our main result is an algorithm that provably estimates all $n$ LOO values for context attribution up to error $\eps$ using $O(\sqrt{\log(1/\eps)}/\eps)$ evaluation calls.
\end{framed}

Existing approaches for context attribution use surrogate models
\citep{cohenwang2024contextcite},
gradient-based attribution scores \citep{qi2024mirage},
and other heuristics \citep{liu2025attribot}.
While fast in practice, such heuristics offer no formal accuracy guarantees for modern AI models and can fail when context pieces interact nonlinearly or when local linearization underestimates deletion effects.
Our context attribution algorithm offers
provable prediction guarantees.
The proof is based on a property of the model called
\emph{stability}.
Similar notions have recently been studied in the setting of
training data attribution~\citep{gunn2026sketch}.
Experimentally, we verify that stability is compatible with modern language models and that our algorithm accurately estimates LOO values on \texttt{Llama-3.2-1B}.

\subsection{Our Contributions}
\label{sec:introduction:our-contributions}

Before informally stating our results,
we introduce the necessary definitions and context.

Let $\bm e_i\in \{0, 1\}^n$ denote the canonical basis vector whose $i$-th entry is 1 and all other entries are 0. We use binary vectors in $\{0, 1\}^n$ to indicate which parts of the context are deactivated ($1$) versus kept active ($0$): thus $\bm 0$ represents the full prompt, while $\bm e_i$ deactivates the $i$-th token.
Relaxing $\{0,1\}^n$ to $[0,1]^n$ naturally encodes ``partial'' deactivation: each downweight $w_i\in [0, 1]$ controls the extent to which the $i$-th token is suppressed.

We formally define a measurement function $\phi: [0, 1]^n\to \R$ as mapping from downweights $\bm w\in [0, 1]^n$ to scalar values.
Our goal is to estimate the set of all LOO values $\phi(\bm e_1), \dots, \phi(\bm e_n)$.

\paragraph{Stability.}
Let $\otimes$ denote the tensor product.
Consider the Taylor expansion of $\phi$ around the origin $\bm0$,
\[
    \phi(\bm w)=\sum_{r\ge0}T_\phi^{(r)}[\bm w^{\otimes r}],
    \qquad
    T_\phi^{(r)}\coloneqq\frac1{r!}D^r \phi(\bm0).
\]
Here $T_\phi^{(r)}$ is the degree-$r$ Taylor coefficient tensor.
The main structural property upon which our guarantees are
based is \emph{stability}
(\Cref{def:stability}).
We say that $\phi$ is $\alpha$-stable if, for every $B>0$
and every integer $r\ge1$, its Taylor coefficients satisfy
\begin{align}
    \norm{T_\phi^{(r)}}_F
    \le \frac{\alpha(B)}{B^r}\,.
    \label{eq:introduction:stability-informal}
\end{align}
In our informal discussion below, we say simply that $\phi$ is \emph{stable} when it is $\alpha$-stable with $\alpha(B) = O(1)$ independent of the context length $n$ and degree $r$.
\paragraph{Our Contributions.} We are now ready to informally state our results.
\begin{theorem}[Informal; see \Cref{thm:optimal-context-attribution,thm:optimal-context-attribution:oracle-lower-bound}]
    For sufficiently small $\eps>0$,
    there is a randomized algorithm for estimating LOO values with the following guarantees.
    For any evaluation function $\phi$ mapping $n$ downweights to a scalar metric,
    \begin{itemize}
        \item the algorithm makes $O(\min\{n,\sqrt{\log(\nicefrac1\eps)}/\eps\})$ evaluation calls;
        \item if $\phi$ is stable, then with probability $0.99$,
            all estimates have error at most $O(\eps)$;
        \item any algorithm achieving error $O(\eps)$ for stable $\phi$ requires $\Omega(\min\{n, \sqrt{\log(\nicefrac1\eps)}/\eps\})$ evaluation calls.
    \end{itemize}
\end{theorem}

The notion of stability was recently introduced by
\citet{gunn2026sketch} for the training data attribution problem,
where the goal is to predict how removing training examples
changes the output of a learning algorithm.
Roughly speaking,
they are able to estimate $k$ arbitrary subset deletion queries up to error $\eps$ in the data attribution setting using $\widetilde O(\log(k)/\eps^2)$ training runs.
Their approach is based on a sketching algorithm for stable functions via forward differentiation.
Our techniques are based on a refinement of their sketch that reduces the variance of the estimator.

\paragraph{Technical Contributions.}
There is a natural adaptation of \citeauthor{gunn2026sketch}'s forward differentiation scheme to our setting
using $\widetilde O(\log(n)/\eps^2)$ evaluation calls. 
Instead of directly sketching $\phi$,
our key observation is to apply forward differentiation to the \emph{gradient circuit} $\grad \phi: [0, 1]^n \to \R^n$ obtained via backpropagation.
By classical backpropagation complexity~\citep{Lin76,BS83,RHW86}, sketching $\grad \phi$ incurs only a constant-factor overhead over sketching $\phi$.
As it turns out, this significantly reduces the variance of the resulting estimators.
Moreover, this only incurs constant overhead when computing $\grad \phi$ using backpropagation and allows us to improve the number of evaluation calls by a factor of $\log(n)/\eps$.
Interestingly, the number of sketches does not depend on the number of tokens $n$!

\subsection{Related Works}
\label{sec:introduction:related-works}

\paragraph{Context Attribution.}
Context attribution studies how input passages affect a model's response.
Existing methods use sparse surrogates fitted to context
ablations~\citep{cohenwang2024contextcite}, gradient
saliency~\citep{qi2024mirage}, and efficient LOO estimation through
KV caching, hierarchical attribution, and proxy
models~\citep{liu2025attribot}.
Related work evaluates generated citations~\citep{gao2023alce}
and improves them using context-ablation
rewards~\citep{chuang2025selfcite}.
We study the number of calls needed to provably approximate all LOO
values simultaneously under stability.

\paragraph{Training Data Attribution.}
Training data attribution studies how training examples affect model
predictions.
Approaches include influence functions and training-trajectory
scores~\citep{koh2017influence,pruthi2020tracin}, learned surrogates
and projected gradients~\citep{ilyas2022datamodels,park2023trak},
and differentiation through
training~\citep{ilyas2025magic,chen2026sketched}.
Accuracy guarantees and corrections for influence-function and
Newton-step approximations have also been studied for convex
learning problems~\citep{rubinstein2025rescaled,rubinstein2026accuracy}.
On a technical level, the closest work to ours is the sketching-based data attribution algorithm due to \citet{gunn2026sketch}.
Our algorithm is based on a refinement of their approach adapted to the context attribution setting.

\paragraph{Feature Attribution.}
Feature attribution assigns importance scores to input features to explain a model's prediction.
Methods include representation
erasure~\citep{li2016understanding}, local surrogate
explanations~\citep{ribeiro2016lime}, and Shapley-value
methods~\citep{lundberg2017shap}.
\citet{kang2024learning} give guarantees for recovering the
M\"obius coefficients of a masking function under a model with
sparse, randomly located interactions.
Our LOO target follows the erasure perspective by measuring the
effect of a single deletion from the full input.
Shapley values instead average marginal contributions across subsets.

\section{Technical Overview}
\label{sec:technical-overview}
We now give a high-level overview of our context attribution scheme.

\paragraph{Notation.}
We write $[n]\coloneqq\{1,\ldots,n\}$.
Let
\[
    T_f^{(r)}\coloneqq\frac1{r!}D^r f(\bm0)
\]
denote the degree-$r$ Taylor coefficient tensor of $f$ at the origin,
and omit the subscript $f$ when the function is clear.
We write $\bm x^{\otimes r}$ for the $r$-fold tensor product
and $T^{(r)}[\bm x_1\otimes\cdots\otimes\bm x_r]$
for contraction against these vectors, without complex conjugation.
The notation $T^{(r)}[\bm x,\cdot]$ contracts only the first index.
We use $\norm{\cdot}_F$ for the Frobenius norm,
the Euclidean norm of the tensor entries.
A dagger denotes conjugate transpose, and
$\iprod{\bm x,\bm y}\coloneqq\bm x^\dagger\bm y$.

\paragraph{Setting.}
We model the evaluation function $\phi$ as an arithmetic circuit over complex numbers (\Cref{def:arithmetic-circuit}).
This is a special case of the computation graphs used in deep learning~\citep{Bau74,GBC16}.
Recall that evaluating $\phi$ takes $O(\size(\phi))$ arithmetic operations.
We note that all of our results translate to modern language models used in practice and this abstraction is mainly to avoid cumbersome technical details.
Indeed, our method only requires that the gates in the computation graph have local polynomial approximations around real inputs and we easily implemented our algorithm on \texttt{Llama-3.2-1B}.

We also note that while the notion of downweights is natural for training algorithms such as stochastic gradient descent,
\eg{}, we downweight the gradient of the $i$-th training example by $w_i$,
there is no consensus on the specific implementation of downweights in the context attribution setting.
This is important as not all implementations will satisfy the stability property that we require.
We implement the downweights through continuous masking of attention scores and observe that this seems compatible with transformer-based models.
See \Cref{sec:continuous-masking} for details.

\subsection{Sketching Arithmetic Circuits.}
\label{sec:technical-overview:sketching-arithmetic-circuits}
Before overviewing our algorithm, it is instructive to abstractly recall \citet{gunn2026sketch}'s sketch for stable arithmetic circuits $f$.

Suppose $f$ is stable in the sense that its Taylor coefficients decay as $\norm{T_f^{(r)}}_F\leq \frac{\alpha(B)}{B^r}$.
We use this property in two ways:
\begin{enumerate}[(1)]
    \item Bound the error from truncating the Taylor expansion;
    \item Bound the variance of sketching the truncated Taylor expansion.
\end{enumerate}
Indeed, assuming stability holds for $B > 2$, truncating the Taylor expansion at order $N=O(\log(\nicefrac1\eps))$ introduces at most error $O(\eps)$.
However, directly computing the exact Taylor coefficients is expensive even for $r\geq 3$ since the degree-$r$ tensor has $n^r$ entries.
A natural idea is to compute projections of the Taylor coefficient tensors using random vectors $\bm g_1, \dots, \bm g_r\in \mathbb{R}^n$ for $1\leq r\leq N$
\[
    T^{(r)} [\bm g_1\otimes \dots \otimes \bm g_r]\,.
\]

\paragraph{Sketching by Random Complex Projections.} As it turns out,
it suffices to use the tensor power of a single standard \emph{complex} Gaussian vector $\bm g\sim\mathcal{CN}(0,I_n)$.
Indeed, given sketches of the form $T^{(r)}[\bm g^{\otimes r}]$,
we can obtain unbiased estimates of $T^{(r)}[\bm w^{\otimes r}]$ as follows.
\begin{align*}
    \E_{\bm g}\left[
        \frac{T^{(r)}[\bm g^{\otimes r}]
        \cdot\iprod{\bm g^{\otimes r},\bm w^{\otimes r}}}{r!}
    \right]
    &= T^{(r)}
      \E_{\bm g}\left[
        \frac{\bm g^{\otimes r}(\bm g^{\otimes r})^\dagger}{r!}
      \right]\bm w^{\otimes r}
    = T^{(r)}[\bm w^{\otimes r}]\,.
\end{align*}
One can bound the variance of this sketch as
\[
    \Var_{\bm g}\left[
      \Re
      \frac{T^{(r)}[\bm g^{\otimes r}]
            (\bm g^\dagger\bm w)^r}{r!}
    \right]
    \le
      \binom{2r}{r}\,\norm{T^{(r)}}_F^2\,\norm{\bm w}_2^{2r}\,.
\]
Let us now focus on the LOO estimates
$f(\bm e_1),\dots,f(\bm e_n)$.
We will estimate the degree-$r$ term of the Taylor expansion of $f(\bm e_i)$ for all $r\in [N]$ and then combine the per-degree estimates to an estimate of $f(\bm e_i)$.
Fix a degree $r\in[N]$.
For independent $\bm g_1,\dots,\bm g_m\sim\mathcal{CN}(0,I_n)$,
define
\[
    Z_{i, r}^{(j)}\coloneqq
    \Re\frac{
      T^{(r)}[\bm g_j^{\otimes r}]
      (\bm g_j^\dagger\bm e_i)^r
    }{r!}\,,\qquad\qquad
    Z_{i, r}\coloneqq
    \frac1m\sum_{j=1}^m
    Z_{i, r}^{(j)}\,.
\]
Then
\[
    \E_{\bm g_1,\dots,\bm g_m}[Z_{i, r}]
    = \E_{\bm g_1}[Z_{i, r}^{(1)}]
    = T^{(r)}[\bm e_i^{\otimes r}]\,,\qquad\qquad
    \Var_{\bm g_1,\dots,\bm g_m}(Z_{i, r})
    =  \frac1m \Var_{\bm g_1}(Z_{i, r}^{(1)})
    \leq \frac{4^r}{m}\norm{T^{(r)}}_F^2\,.
\]
By Chebyshev's inequality, with contant probability,
the error at degree $r$ is at most the square root of the variance.
Unfortunately, we cannot directly employ $Z_{i, r}$ as we do not have sufficiently strong concentration of the sample average,
so we must employ a mean estimation subroutine for heavy-tailed distributions.
For instance, using a trimmed mean estimator $\Trim(Z_{i, r}^{(1)}, \dots, Z_{i, r}^{(m)})$ which removes a few of the largest and smallest samples~\citep{OOR25},
we can show that the estimation error is at most
\[
    O\left(
      2^r\norm{T^{(r)}}_F
      \sqrt{\frac{\log(\nicefrac{nN}\delta)}m}
    \right)
\]
simultaneously for all LOO queries and degrees $r\in [N]$,
with probability at least $1-\delta$.
Assuming stability for $B > 2$, we thus have
\[
    \abs{\sum_{r=1}^N T^{(r)}[\bm e_i^{\otimes r}] - \sum_{r=1}^N \Trim(Z_{i, r}^{(1)}, \dots, Z_{i, r}^{(m)})}
    = O\left(
        \sqrt{\frac{\log(\nicefrac{nN}\delta)}{m}}
      \right)\,.
\]
Thus $m=O(\log(\nicefrac{nN}\delta)/\eps^2)$ sketches suffice to estimate all LOO values up to error $\eps$.

\paragraph{Sketching by Forward Differentiation.} Fix $\bm g\in \mathbb{C}^n$.
A key observation is that we can compute these directional derivatives $T^{(r)}[\bm g^{\otimes r}]$ for all $r\in [N]$ simultaneously using forward differentiation in the direction $\bm g$ using $O(\size(f)N\log N)$ operations~\citep{Wen64}.
This can be interpreted as evaluating $f(z \bm g)$ over the polynomial ring $\C[z]/(z^{N+1})$.
All in all, the sketch can be efficiently computed in $O(\log(\nicefrac{nN}\delta)/\eps^2)$ evaluations and each evaluation takes time $\widetilde O(\size(f) \log(\nicefrac1\eps))$.

\subsection{Our Approach}
\label{sec:technical-overview:our-approach}
Our algorithms are driven by a refinement of the forward sketches for stable arithmetic circuits.

\paragraph{Backward--Forward Sketches.}
The counterpart to forward differentiation is backward differentiation~\citep{Lin76,RHW86}.
Roughly speaking, backpropagation constructs an auxiliary circuit of size $O(\size(f))$ that computes
\[
    \bm w\mapsto (f(\bm w), \grad f(\bm w))\,.
\]
The advantage of backpropagation is that it computes the exact first-order gradient $\grad f(\bm w)$.
However, there is no higher-order analogue in the sense that there is no known $\size(f)\cdot\poly(r)$-time algorithm for computing all degree-$r$ partial derivatives.

Our key observation is that we can perform forward differentiation on the backpropagation circuit itself!
In Taylor coefficients,
this corresponds to sketches of the form
\[
    T^{(r)}[\bm e_i\otimes \bm g^{\otimes r-1}]\,, i\in [n]\,.
\]
In other words,
instead of randomly sketching the entire tensor $T^{(r)}$,
we sketch disjoint slices $T^{(r)}[\bm e_i, \cdot]$.

Using our new sketches, we can compute LOO estimates in an almost identical fashion.
The constant and linear terms are available exactly,
so fix a degree $2\le r\le N$.
As the $\bm e_i$'s are orthonormal,
our estimates simplify to
\begin{align*}
    \bar Z_{i,r}^{(j)}
    &\coloneqq
    \Re\sum_{\ell\in[n]}
    \frac{
      T^{(r)}[\bm e_\ell\otimes\bm g_j^{\otimes(r-1)}]
      (\bm e_\ell^\top\bm e_i)
      (\bm g_j^\dagger\bm e_i)^{r-1}
    }{(r-1)!}\\
    &=
    \Re\frac{
      T^{(r)}[\bm e_i\otimes\bm g_j^{\otimes(r-1)}]
      (\bm g_j^\dagger\bm e_i)^{r-1}
    }{(r-1)!},\\
    \bar Z_{i,r}
    &\coloneqq
    \frac1m\sum_{j=1}^m\bar Z_{i,r}^{(j)},\\
    \E_{\bm g_1,\dots,\bm g_m}[\bar Z_{i,r}]
    &=
    \E_{\bm g_1}[\bar Z_{i,r}^{(1)}]
    =T^{(r)}[\bm e_i^{\otimes r}]\,.
\end{align*}

\begin{example}\label{ex:technical-overview:backward-forward-sketches}
    Let us illustrate our estimator on $T^{(2)}$ for the sake of clarity.
    Sketching the full tensor and estimating $T^{(2)}[\bm w^{\otimes 2}]$ corresponds to
    \[
                  \frac12(\bm g^\top T^{(2)}\bm g)
          (\bm g^\dagger\bm w)^2
        \,.
    \]
    On the other hand,
    sketching the slices and then estimating corresponds to
    \[
        \bm w^\top T^{(2)} \bm g (\bm g^\dagger \bm w)\,.
    \]
    Intuitively, 
    this reduces the amount of randomness in the estimator.
\end{example}

Let us formalize the intuition in \Cref{ex:technical-overview:backward-forward-sketches}.
Define
\[
    v_{i, r}\coloneqq \binom{2r-2}{r-1}
    \norm{T^{(r)}[\bm e_i,\cdot]}_F^2\,.
\]
By the orthogonality of $\bm e_i$'s and a Gaussian moment bound on $\bar Z_{i, r}$, 
we have that
\[
    \Var_{\bm g_1,\dots,\bm g_m}(\bar Z_{i,r})
    =
    \frac1m\Var_{\bm g_1}(\bar Z_{i,r}^{(1)})
    \le
    \frac1m v_{i, r}\,.
\]
Here $v_{i, r}$ is an upper bound for the degree-$r$ variance.
Similar to Gunn's approach, we will apply a separate trimmed mean
$\Trim(\bar Z_{i,r}^{(1)},\dots,\bar Z_{i,r}^{(m)})$
for every $i$ and $r$.
The degree-$r$ error is
\[
    O\left(
      \sqrt{\frac{v_{i,r}\log(\nicefrac{nN}\delta)}m}
    \right).
\]

\paragraph{Variance of the Estimator.}
We then add the degree-wise estimates and their error bounds.
Define
\[
    s_i\coloneqq\sum_{r=2}^{N}\sqrt{v_{i,r}}\,.
\]
Following our per-degree calculation above,
the total error for estimating $f(\bm e_i)$ from samples is
\[
    \abs{
      \sum_{r=2}^N T^{(r)}[\bm e_i^{\otimes r}]
      -\sum_{r=2}^N
        \Trim(\bar Z_{i,r}^{(1)},\dots,\bar Z_{i,r}^{(m)})
    }
    = O\left(
      s_i\sqrt{\frac{\log(\nicefrac{nN}\delta)}m}
    \right)\,.
\]
Ignoring the common $\Ot(\sqrt{\nicefrac1m})$ term,
it follows that the root-mean-square estimation error across LOO queries scales as
\begin{align*}
    \sqrt{\sum_i s_i^2}
    &= \sqrt{\sum_i \left( \sum_{r=2}^N \sqrt{v_{i, r}} \right)^2}
    \leq \sum_{r=2}^{N} \sqrt{\sum_i v_{i, r}} &&\text{Minkowski's inequality} \\
    &=
    \sum_{r=2}^{N}
      \sqrt{\binom{2r-2}{r-1}\,
      \sum_i \norm{T^{(r)}[\bm e_i, \cdot]}_F^2} \\
    &\leq \sum_{r=2}^{N}
      \sqrt{4^{r-1}\,
      \norm{T^{(r)}}_F^2}
    \leq \sum_{r=2}^N 2^r \norm{T^{(r)}}_F\\
    &= O(1)\,.
\end{align*}
Compared to vanilla forward differentiation of $f$,
we note that the total sum of errors from backward-forward sketches scales with the tensor norm,
compared to only the maximum error of the former.

By Markov's inequality,
at most $O(\nicefrac1\eps)$ of the squared error $s_i^2$'s
are more than $O(\eps)$.
We select the $O(\nicefrac1\eps)$ largest $s_i$'s and evaluate the corresponding values
$f(\bm e_i)$ exactly in $O(\size(f)/\eps)$ time.
Every remaining query has $s_i=O(\sqrt\eps)$,
so its total sampling error is
\[
    O\left(
      \sqrt{\frac{\eps\log(\nicefrac{nN}\delta)}m}
    \right).
\]
Thus $m=O(\log(\nicefrac{nN}\delta)/\eps)$ sketches suffice
for $\eps$-error.

In order to estimate $s_i$, 
we can estimate the $v_{i,r}$ using a trimmed mean estimator $\widehat v_{i,r}$ from samples.
Using a fourth-moment bound,
we can achieve constant relative error estimates using
$O(4^{N-1}\log(\nicefrac{nN}\delta))$ sketches.
For $N=\log_4(\nicefrac1\eps)$,
this is $O(\log(\nicefrac{nN}\delta)/\eps)$.
We can thus estimate $s_i$ up to constant relative error
\[
    \widehat s_i\coloneqq\sum_{r=2}^{N}\sqrt{\widehat v_{i,r}}\,.
\]

All in all,
the illustrated approach requires $O(\log(\nicefrac{nN}\delta)/\eps)$ model calls with each call taking $\widetilde O(\size(f)\log(\nicefrac1\eps))$ operations.

\paragraph{Sharper Bounds.}
The factor $\log n$ above comes from taking a union bound over all coordinates. 
In \Cref{sec:optimal-context-attribution:our-attribution-scheme},
we argue that we only need this union bound for $O(\nicefrac1\eps)$ coordinates with large
error proxies.
This reduces the number of queries to $O(\log(\nicefrac{N}{\eps\delta})/\eps)$.  
Moreover, we used a simple upper bound $\binom{2r-2}{r-1}\leq 4^{r-1}$ in controlling the sum of errors.
We will see that a tighter bound combined with some more careful choice of hyperparameters allows us to further improve the query complexity to $O((\sqrt{\log(\nicefrac{N}{\eps})} + \log(\nicefrac1\delta))/\eps)$.

\subsection{Query Complexity Lower Bound}
\label{sec:technical-overview:query-complexity-lower-bound}

We highlight the key reduction in the proof of the oracle lower bound
proved in \Cref{sec:optimal-context-attribution:lower-bound}.

\paragraph{Oracle Model.}
We assume that we are given access to a stable function $f: [0, 1]^n\to \R$ in the form of an oracle.
For any input $\bm g\in \C^n$, and integer $r \geq 1$,
the oracle provides
\[
    (\bm w, \bm g, r)\mapsto \left( f(\bm w), \grad f(\bm w), D_{\bm g}^1\grad f(\bm w), \dots, D_{\bm g}^r\grad f(\bm w) \right)\,.
\]
Thus it provides exact zero-th and first-order information at $\bm w$,
in addition to directional derivatives of arbitrary order,
\[
    D_{\bm g}^k \grad f(\bm w) \coloneqq \frac{\partial^k}{\partial z^k} \eval{\grad f(\bm w + z\bm g)}_{z=0}\,,
\] 
exactly as our algorithm requires.
As we will see, the terms $r > 1$ will not be useful for the lower bound instance.

\paragraph{Diagonal Estimation Lower Bound.}
We show that our algorithm uses the optimal $\Theta(\sqrt{\log(\nicefrac1\eps)}/\eps)$ queries in the oracle model above.
It suffices to consider quadratic functions
$f(\bm w)=\bm w^\top A\bm w$ for a random symmetric matrix $A\in \R^{n\times n}$.
Note that $f$ is stable since its higher-order Taylor coefficients are all zero.
Observe that
\begin{align*}
    f(\bm w)&=\bm w^\top A\bm w\,, \\
    \grad f(\bm w)&=2A\bm w\,,\\
    D_{\bm g}^1 \grad f(\bm w) &= 2A\bm g\,,\\
    D_{\bm g}^r \grad f(\bm w) &= 0\,,\quad \forall r > 1\,.
\end{align*}
Hence the oracle essentially returns two matrix-vector products $A\bm w, A\bm g$ in every query.

Since $\bm e_i^\top A \bm e_i = A_{ii}$,
estimating the leave-one-out values in this oracle model essentially amounts to estimating
the diagonal of $A$ from matrix--vector products.
We adapt the lower bound techniques for this problem due to \citet*{amsel2024fixed}.
The main change is to measure the maximum estimation error,
whereas their argument is based on Frobenius error.

\section{Preliminaries}
\label{sec:preliminaries}
In this section, we establish the notation use throughout the rest of this work and review the necessary background.

\subsection{Notation}
\label{sec:preliminaries:notation}

For a positive integer $n$, write $[n]\coloneqq\{1,\ldots,n\}$.
Let $\bm e_i$ denote the $i$-th standard basis vector, and let
$\bm1_D\in\R^n$ denote the indicator vector of $D\subseteq[n]$.
For a complex vector $\bm x$, write $\overline{\bm x}$ for its complex
conjugate and $\bm x^\dagger\coloneqq\overline{\bm x}^{\top}$ for its
conjugate transpose. All logarithms are natural unless a base is specified.

\paragraph{Tensors And Inner Products.}
For tensors $S,T\in(\C^n)^{\otimes r}$ with $r\ge1$, we use the Frobenius inner product
\[
    \langle S,T\rangle\coloneqq
    \sum_{j_1,\ldots,j_r=1}^n
      \overline{S_{j_1,\ldots,j_r}}T_{j_1,\ldots,j_r},
    \qquad
    \norm{T}_F^2\coloneqq\langle T,T\rangle,
\]
which is conjugate-linear in its first argument.
For vectors, we use the Euclidean norm $\norm{\bm x}_2$.
For a (complex) tensor $T\in(\C^n)^{\otimes r}$ and vectors
$\bm x_1,\ldots,\bm x_r\in\C^n$, write
\[
    T[\bm x_1\otimes\cdots\otimes\bm x_r]
    \coloneqq
    \sum_{j_1,\ldots,j_r=1}^n
    T_{j_1,\ldots,j_r}
      (\bm x_1)_{j_1}\cdots(\bm x_r)_{j_r}\in\C,
\]
and abbreviate this as $T[\bm x^{\otimes r}]$ when all arguments equal $\bm x$.
Note that contraction does not conjugate the arguments.
We write $T[\bm x,\cdot]\in(\C^n)^{\otimes(r-1)}$ for the partial contraction that retains $r-1$ indices,
and $T[\cdot,\bm x^{\otimes(r-1)}]\in\C^n$ for the partial contraction that retains a single index.
We identify order-zero tensors with scalars.

Let $\Sym^r(\C^n)$ denote the subspace of
$(\C^n)^{\otimes r}\cong\C^{n^r}$ consisting of tensors invariant
under permutations of their indices, and let
$\Pi_{\mathrm{sym}}^{(n,r)}$ denote its orthogonal projector.

\paragraph{Taylor Coefficients.}
For a function $f:\C^n\to\C$ that is holomorphic in a neighborhood
of the origin, define its normalized Taylor tensors by
\[
    T_f^{(r)}\coloneqq\frac1{r!}D^r f(\bm0).
\]
When the function is clear, we omit the subscript $f$.
For $r\ge1$, the tensor $T^{(r)}$ belongs to $\Sym^r(\C^n)$.
The degree-zero term $T^{(0)}=f(\bm0)$ is a complex scalar, and
$T^{(1)}=\grad f(\bm0)$ is a complex vector.

\paragraph{Complex Gaussians.}
Let $\bm a,\bm b$ be independent $\mathcal N(0,I_n)$ vectors.
Let
\[
    \bm g\coloneqq\frac{\bm a+\ii\bm b}{\sqrt2}
\]
denote a standard complex Gaussian vector and write
$\bm g\sim\mathcal{CN}(0,I_n)$. Note that
$\E_{\bm g}[g_i\overline{g_j}]$ is $1$ when $i=j$ and $0$ otherwise,
and $\E_{\bm g}[g_ig_j]=0$ for all $i,j$.
For a complex random variable $X$, write
$\Var_X(X)\coloneqq\E_X|X-\E_X X|^2$.

\subsection{Arithmetic Circuits and Computation}
\label{sec:preliminaries:arithmetic-circuits-and-computation}

\paragraph{Computation.}
We count complex arithmetic operations as constant time and
assume that complex numbers can be stored exactly in constant
space. We further assume random access to memory with a word
size large enough for each lookup to take constant time.

\paragraph{Arithmetic Circuits.}

We formally represent a neural network as an \emph{arithmetic circuit} over complex numbers.
This is a special case of the computation graphs used in deep learning~\citep{Bau74,GBC16}.

\begin{definition}[Arithmetic Circuit]
\label[definition]{def:arithmetic-circuit}
An arithmetic circuit $f:\C^n\to\C^p$ is a finite directed acyclic
graph with labeled vertices. Each vertex with in-degree zero is
labeled by an input index $i\in[n]$ or a complex constant.
Each vertex with positive in-degree is labeled by $+$ or $\times$
and computes the sum or product of its incoming values. Its value
is passed to all of its children. We do not restrict the in-degree
or out-degree. Each vertex with out-degree zero is also labeled
by an output index in $[p]$. Input indices may be reused, and each
output index labels exactly one vertex.
For an input $\bm w\in\C^n$, a vertex labeled by input index $i$
receives $w_i$. Evaluating the gates in topological order gives
$f(\bm w)$ in output-index order.
We define its size, denoted $\size(f)$, as the maximum of $n$,
$p$, and the number of edges.
\end{definition}

A scalar-valued arithmetic circuit with constants computes a
polynomial. Its Taylor expansion is therefore finite and exact,
including at complex inputs:
\[
    f(\bm w)=\sum_{r\ge0}T^{(r)}[\bm w^{\otimes r}],
    \qquad \bm w\in\C^n,
\]
where $T^{(r)}=0$ above the degree of the polynomial.

We work under \Cref{def:arithmetic-circuit} to avoid cumbersome
technical details. Gates such as ReLU, GELU, and Softmax may
be replaced by chosen polynomial approximations before applying
the results. 
Indeed, we easily implemented the procedure on the open-weight \texttt{Llama-3.2-1B} model.

\section{The Toolkit}
\label{sec:the-toolkit}
We now present our algorithm design tool kit to obtain our context attribution algorithm.
In \Cref{sec:the-toolkit:gaussian-sketches-of-symmetric-tensors}, 
we establish some basic property of complex Gaussian sketches.
Then in \Cref{sec:the-toolkit:backward-and-forward-differentiation},
we describe our main sketching subroutine which involves both backpropagation and forward differentiation.
Next, we formally define stability and explore basic intuition and consequences of this definition in \Cref{sec:the-toolkit:stability}.
Finally, we state the mean estimation subroutine we employ for potentially heavy-tailed distributions in \Cref{sec:the-toolkit:trimmed-mean-estimation}.

\subsection{Gaussian Sketches Of Symmetric Tensors}
\label{sec:the-toolkit:gaussian-sketches-of-symmetric-tensors}

Let $\bm g$ be a standard complex Gaussian vector.
The following proposition explains why contractions against
$\bm g^{\otimes r}$ suffice to sketch symmetric tensors.
We note that the same fact does not hold for real Gaussian vectors and it is necessary to use \emph{complex} Gaussians.

\begin{proposition}[Complex Gaussian Moments; {\cite[Eq.~(9)]{Har13}}]
\label[proposition]{prop:the-toolkit:complex-gaussian-moment-identity}
Let $\bm g\sim\mathcal{CN}(0,I_n)$.
The following hold for every integer $r\ge1$.
\begin{enumerate}[(i)]
    \item \label{item:the-toolkit:complex-gaussian-moment-identity-proposition:moment-operator}
    The Gaussian moment operator satisfies
    \[
        \E_{\bm g}\!\left[
          \bm g^{\otimes r}(\bm g^{\otimes r})^\dagger
        \right]
        =r!\,\Pi_{\mathrm{sym}}^{(n,r)}\,.
    \]
    \item \label{item:the-toolkit:complex-gaussian-moment-identity-proposition:second-moment}
    For every tensor $A\in(\C^n)^{\otimes r}$,
    \[
        \E_{\bm g}|A[\bm g^{\otimes r}]|^2
        =r!\norm{\Pi_{\mathrm{sym}}^{(n,r)}A}_F^2
        \le r!\norm{A}_F^2.
    \]
\end{enumerate}
\end{proposition}

The next lemma records the moments needed by our estimator.
Its first conclusion will let us estimate the size of a tensor
contraction, and its second conclusion gives an unbiased estimate
of the contraction in a prescribed direction and bounds its
second moment.

\begin{lemma}[Gaussian Tensor Moments]
\label[lemma]{lem:the-toolkit:gaussian-tensor-moments}
Let $S\in\Sym^r(\R^n)$, $\bm x\in\R^n$, and
$\bm g\sim\mathcal{CN}(0,I_n)$, where $r\ge1$. The following hold.
\begin{enumerate}[(i)]
    \item \label{item:the-toolkit:gaussian-tensor-moments-lemma:energy-moments}
    The second and fourth moments satisfy
    \[
        \E_{\bm g}|S[\bm g^{\otimes r}]|^2=r!\norm{S}_F^2,
        \qquad
        \E_{\bm g}|S[\bm g^{\otimes r}]|^4
        \le(2r)!\norm{S}_F^4.
    \]
    \item \label{item:the-toolkit:gaussian-tensor-moments-lemma:sketch-moments}
    The directional sketch satisfies
    \[
        \E_{\bm g}\!\left[
          \frac{S[\bm g^{\otimes r}](\bm g^\dagger\bm x)^r}{r!}
        \right]=S[\bm x^{\otimes r}]
    \]
    and
    \[
        \E_{\bm g}\left|
          \frac{S[\bm g^{\otimes r}](\bm g^\dagger\bm x)^r}{r!}
        \right|^2
        \le\binom{2r}{r}\norm{S}_F^2\norm{\bm x}_2^{2r}.
    \]
\end{enumerate}
\end{lemma}

\begin{proof}
For \Cref{item:the-toolkit:gaussian-tensor-moments-lemma:energy-moments}, symmetry of $S$ and
\Cref{item:the-toolkit:complex-gaussian-moment-identity-proposition:second-moment} give
\[
    \E_{\bm g}|S[\bm g^{\otimes r}]|^2
    =r!\norm{S}_F^2.
\]
Applying \Cref{item:the-toolkit:complex-gaussian-moment-identity-proposition:second-moment} in order $2r$ gives
\begin{align*}
    \E_{\bm g}|S[\bm g^{\otimes r}]|^4
    &=\E_{\bm g}|(S\otimes S)[\bm g^{\otimes 2r}]|^2\\
    &=(2r)!\norm{\Pi_{\mathrm{sym}}^{(n,2r)}
                    (S\otimes S)}_F^2\\
    &\le(2r)!\norm{S}_F^4.
\end{align*}

For \Cref{item:the-toolkit:gaussian-tensor-moments-lemma:sketch-moments}, write
\[
    Y=\frac{S[\bm g^{\otimes r}]
             (\bm g^\dagger\bm x)^r}{r!}.
\]
Since $S$ is real and $\bm x^{\otimes r}$ is symmetric,
\Cref{item:the-toolkit:complex-gaussian-moment-identity-proposition:moment-operator} gives
\begin{align*}
    \E_{\bm g}Y
    &=\frac1{r!}
      \left\langle S,
        \E_{\bm g}\!\left[
          \bm g^{\otimes r}(\bm g^{\otimes r})^\dagger
        \right]\bm x^{\otimes r}
      \right\rangle\\
    &=\langle S,\bm x^{\otimes r}\rangle
     =S[\bm x^{\otimes r}].
\end{align*}
Since $\bm x$ is real, \Cref{item:the-toolkit:complex-gaussian-moment-identity-proposition:second-moment} in order $2r$
gives
\begin{align*}
    \E_{\bm g}|Y|^2
    &=\frac1{(r!)^2}\E_{\bm g}\left|
        (S\otimes\bm x^{\otimes r})
        [\bm g^{\otimes 2r}]
      \right|^2\\
    &=\binom{2r}{r}
      \norm{\Pi_{\mathrm{sym}}^{(n,2r)}
        (S\otimes\bm x^{\otimes r})}_F^2\\
    &\le\binom{2r}{r}\norm{S}_F^2\norm{\bm x}_2^{2r}.
\end{align*}

\end{proof}

\subsection{Backward And Forward Differentiation}
\label{sec:the-toolkit:backward-and-forward-differentiation}
The standard method for computing the gradient of a scalar-valued arithmetic circuit $f$ is backward differentiation (backpropagation)~\citep{RHW86}.
While it computes the full first-order gradient vector exactly in $O(\size(f))$ time,
explicitly computing or storing the full order-$r$ derivative tensor
requires $\binom{n+r-1}{r}$ entries in general,
even after accounting for symmetry.
\begin{proposition}[Backward Differentiation;
{\cite{RHW86}, \cite[Theorem~9.10]{CKW11}}]
\label[proposition]{prop:the-toolkit:backward-differentiation}
Let $f:\C^n\to\C$ be an arithmetic circuit.
Backpropagation constructs an arithmetic circuit of size
$O(\size(f))$ computing
\[
    \bm w\longmapsto\bigl(f(\bm w),\grad f(\bm w)\bigr).
\]
\end{proposition}

On the other hand,
forward differentiation allows us to compute all \emph{directional} derivatives up to order $N$ in time $O(\size(f) N\log N)$.
Roughly speaking,
forward differentiation at point $\bm x$ in direction $\bm g$ is accomplished by evaluating $f(\bm x+z\bm g)$ over the polynomial ring $\C[z]/(z^{N+1})$.
Here $z$ is a formal polynomial variable.
\begin{proposition}[Forward Differentiation;
{\cite{Wen64}, \cite[Section~3.1]{gunn2026sketch}}]
\label[proposition]{prop:the-toolkit:forward-differentiation}
Let $F:\C^n\to\C^p$ be an arithmetic circuit,
$\bm g\in\C^n$ a direction, and $N\ge1$ an integer.
Evaluating $F$ over $\C[z]/(z^{N+1})$ at input $z\bm g$
computes
\[
    [z^r]F(z\bm g)
    =\frac1{r!}D^rF(\bm0)[\bm g^{\otimes r}]
    \;\in\C^p,
    \qquad 0\le r\le N,
\]
in $O(\size(F)\,N\log N)$ arithmetic operations and
$O(\size(F)\,N)$ working storage.
\end{proposition}

Our key observation is that the circuit obtained by backward differentiation is itself an
arithmetic circuit, with only constant-factor overhead.
We can therefore apply forward differentiation to the backpropagation circuit and obtain sliced sketches essentially for free! 
We refer to this procedure as a \emph{backward--forward sketch}.

Specifically, for a direction $\bm g\in\C^n$ and an integer
$N\ge1$, a backward--forward sketch computes the coefficients
of $\grad f(z\bm g)$ through degree $N-1$, giving
\[
    \bm c_r(\bm g)\coloneqq\frac1r[z^{r-1}]\grad f(z\bm g)
    =T^{(r)}[\cdot,\bm g^{\otimes(r-1)}]
    \;\in\C^n\,,
    \qquad 1\le r\le N\,.
\]

By \Cref{prop:the-toolkit:backward-differentiation,prop:the-toolkit:forward-differentiation}, a single sketch
computes these vectors in
$O(\size(f)\,N\log N)$ arithmetic operations and
$O(\size(f)\,N)$ working storage.
We retain $f$ as an output of the gradient circuit, so the
constant coefficients also give $f(\bm0)$ and $\grad f(\bm0)$.

\subsection{Stability}
\label{sec:the-toolkit:stability}

The key structural property
introduced by \citet{gunn2026sketch}
is \emph{stability}.
\begin{definition}[Stability]
\label[definition]{def:stability}
Let $f$ be an arithmetic circuit and let
\mbox{$\alpha:(0,\infty)\to[0,\infty]$}.
We say that $f$ is \emph{$\alpha$-stable} if, for every $B>0$
and every integer $r\ge1$,
the coefficients of the Taylor expansion about the origin satisfy
\[
    \norm{T^{(r)}}_F\le\frac{\alpha(B)}{B^r}\,.
\]
\end{definition}
To see why this might be a natural property, 
recall that a complex-valued function \mbox{$f: \C^n\to \C$} has a \emph{modulus of continuity at $\bm0$} if, for every $\bm x\ne \bm0$,
\[
    |f(\bm x) - f(\bm 0)|\leq \alpha(\norm{\bm x- \bm0}_2)\,.
\] 
If this is the case,
and $f$ is in addition holomorphic,
then Cauchy's estimate~\cite[Theorem~10.26]{Rud87} gives
\[
    \norm{T^{(r)}}_{\mathrm{op}} \leq \frac{\alpha(B)}{B^r}\,.
\]
Here
\[
    \norm{T^{(r)}}_{\mathrm{op}}
    \coloneqq\sup_{\norm{\bm u}_2=1}
      |T^{(r)}[\bm u^{\otimes r}]|,
\]
and the bound holds for every $B>0$ and $r\ge1$.
Stability strengthens this coefficient bound by controlling
the Frobenius norm.

An immediate consequence of \Cref{def:stability} is the following
bound on the Taylor truncation error.
\begin{lemma}[Taylor Truncation]
\label[lemma]{lem:the-toolkit:taylor-truncation}
Suppose $f$ is $\alpha$-stable.
For every nonzero $\bm w\in\C^n$ with
$\alpha(4\norm{\bm w}_2)<\infty$ and every integer $N\ge1$,
\[
    \left|
      f(\bm w)-\sum_{r=0}^{N}
      T^{(r)}[\bm w^{\otimes r}]
    \right|
    \le\frac{4^{-N}}3\alpha(4\norm{\bm w}_2).
\]
\end{lemma}

\begin{proof}
For every $r>N$, stability gives
\[
    |T^{(r)}[\bm w^{\otimes r}]|
    \le\norm{T^{(r)}}_F\norm{\bm w}_2^r
    \le4^{-r}\alpha(4\norm{\bm w}_2).
\]
Sum over $r>N$ and use
$\sum_{r>N}4^{-r}=4^{-N}/3$.
\end{proof}

\subsection{Trimmed Mean Estimation}
\label{sec:the-toolkit:trimmed-mean-estimation}

A key subroutine of our context attribution algorithm is estimating the mean of a random variable given only a second moment bound.
In general, this is insufficient for the exponential concentration
needed for logarithmic dependence on the inverse failure probability $\nicefrac1\delta$.
We therefore use the trimmed mean, a classical mean estimator
for possibly heavy-tailed observations.
In 1-D, we remove the $k$ smallest and largest observations
and take the average of the remaining observations.

\begin{definition}[Trimmed Mean; {\cite[Section~2.3]{LM19}}]
\label[definition]{def:trimmed-mean}
For real numbers $y_1,\ldots,y_m$, write
$y_{(1)}\le\cdots\le y_{(m)}$ for their nondecreasing rearrangement.
For an integer $0\le k<m/2$, define
\[
    \Trim_k(y_1,\ldots,y_m)
    \coloneqq\frac1{m-2k}\sum_{j=k+1}^{m-k}y_{(j)}.
\]
\end{definition}

The next proposition bounds the error of the trimmed mean under
a finite-variance assumption.

\begin{proposition}[Trimmed-Mean Concentration; {\cite[Theorem~5.1.1]{OOR25}}]
\label[proposition]{prop:context-attribution:trimmed-mean-concentration}
Let $X_1,\ldots,X_m$ be independent, identically distributed real
random variables with mean $\mu$ and finite variance $\sigma^2$.
Fix $\delta\in(0,1)$ and suppose $m\ge64\log(\nicefrac{16}\delta)$.
Then, with probability at least $1-\delta$ over
$X_1,\ldots,X_m$,
\[
    \left|
      \Trim_{\log(\nicefrac{16}\delta)}
      (X_1,\ldots,X_m)-\mu
    \right|
    \le7\sigma\sqrt{\frac{\log(\nicefrac{16}\delta)}m}\,.
\]
\end{proposition}

\section{Context Attribution}
\label{sec:optimal-context-attribution}

In this section, we study the context attribution problem (\Cref{sec:optimal-context-attribution:the-context-attribution-problem}) and present our algorithm for estimating LOO values (\Cref{sec:optimal-context-attribution:our-attribution-scheme}).

\subsection{The Context Attribution Problem}
\label{sec:optimal-context-attribution:the-context-attribution-problem}
Before presenting our main result,
we remind the reader of the setting of the context attribution problem.

\paragraph{The Setting.}
Let $\phi:\C^n\to\C$ be an arithmetic circuit that is real-valued
on $[0,1]^n$. Its input consists of downweights
$\bm w=(w_1,\ldots,w_n)$, where $w_i=0$ includes the $i$-th token
and $w_i=1$ deletes it.
Our goal is to estimate all LOO values
$\phi(\bm e_1),\ldots,\phi(\bm e_n)$.
For transformer-based language models,
\Cref{sec:continuous-masking} describes an implementation of
these downweights using continuous attention masks.
The algorithm computes backward--forward sketches of $\phi$
and evaluates a few high variance LOO values exactly.

\subsection{Our Attribution Algorithm}
\label{sec:optimal-context-attribution:our-attribution-scheme}
We now present our main result for estimating LOO values.

\begin{theorem}[LOO Attribution]
\label{thm:optimal-context-attribution}
Let $\phi:\C^n\to\C$ be an arithmetic circuit that is real-valued
on $[0,1]^n$, and fix $\eps,\delta\in(0,1)$.
There is a randomized algorithm for estimating all LOO values
with the following guarantees.
\begin{enumerate}[(i)]
    \item \label{item:optimal-context-attribution:complexity}
    The algorithm takes
        \[
        \widetilde O\left(
          \frac{\size(\phi)}{\eps}
          \left(\sqrt{\log(\nicefrac1\eps)}+\log(\nicefrac1\delta)\right)
        \right)
    \]
    arithmetic operations and uses
        \[
        \widetilde O\left(
          \size(\phi)+\frac n\eps
          \left(\sqrt{\log(\nicefrac1\eps)}+\log(\nicefrac1\delta)\right)
        \right)
    \]
    auxiliary space.

    \item \label{item:optimal-context-attribution:accuracy}
    If $\phi$ is $\alpha$-stable and $\alpha(4)<\infty$, then,
    with probability at least $1-\delta$ over the Gaussian
    directions, simultaneously for every $i\in[n]$,
    \[
        |\widehat L_i-\phi(\bm e_i)|
        \le\eps\alpha(4).
    \]

    \item \label{item:optimal-context-attribution:evaluations}
    The total number of evaluation calls is
        \[
        O\left(\min\left\{
          n,\frac{\sqrt{\log(\nicefrac1\eps)}+\log(\nicefrac1\delta)}{\eps}
        \right\}\right).
    \]
\end{enumerate}
\end{theorem}
As previously remarked, our algorithm improves the dependence on
$\nicefrac1\eps$ from quadratic to linear in the number of arithmetic
operations compared with direct forward differentiation sketches
(\Cref{thm:context-attribution:forward-sketch-baseline}).

The rest of this subsection is dedicated to proving \Cref{thm:optimal-context-attribution}.

\paragraph{Degree-Wise Estimates.}
Our strategy is to sketch the degree-$r$ Taylor coefficients separately for each $r\in [N]$ and then add the resulting
estimates and error bounds.

For simplicity, we write $T^{(r)}=T_\phi^{(r)}$.
Recall from \Cref{sec:the-toolkit:backward-and-forward-differentiation} that for a direction $\bm g\in\C^n$ and an integer
$N\ge1$, a backward--forward sketch computes the coefficients
of $\grad\phi(z\bm g)$ through degree $N-1$, giving
\[
    \bm c_r(\bm g)\coloneqq\frac1r[z^{r-1}]\grad\phi(z\bm g)
    =T^{(r)}[\cdot,\bm g^{\otimes(r-1)}]
    \;\in\C^n\,,
    \qquad 1\le r\le N\,.
\]
Moreover, the constant and linear terms are available exactly.

For independent $\bm g_1,\ldots,\bm g_m\sim\mathcal{CN}(0,I_n)$,
define, for every $i\in[n]$ and $2\le r\le N$, the following estimates of the degree-$r$ Taylor expansion along $\bm e_i$
\[
    \bar Z_{i,r}^{(j)}
    \coloneqq
    \Re\frac{
      \bm e_i^\top\bm c_r(\bm g_j)
      (\bm g_j^\dagger\bm e_i)^{r-1}
    }{(r-1)!}\,.
\]
As in \Cref{sec:technical-overview:our-approach}, define a variance
proxy for each degree and an error proxy for their sum:
\[
    v_{i,r}\coloneqq
    \binom{2r-2}{r-1}\norm{T^{(r)}[\bm e_i,\cdot]}_F^2,
    \qquad
    s_i\coloneqq\sum_{r=2}^{N}\sqrt{v_{i,r}}\,.
\]

\paragraph{Algorithm Description.}
We first describe the algorithm for a truncation degree $N$,
$m$ backward--forward sketches, $K$ exact evaluations, and trimmed count $k$.
We choose these parameters in the proof of the theorem.
If $n\le m+K$, we can trivially evaluate all LOO values exactly.
Thus we proceed assuming otherwise.

Our first step is to compute backward--forward sketches of the degree-$r$ Taylor tensors simultaneously for each $r\in [N]$.
For each random complex Gaussian $\bm g\in \C^n$,
this gives us sketches of the form $T^{(r)}[\bm e_i\otimes \bm g^{\otimes r-1}]$ for $i\in [n], r\in [N]$.
For each LOO value $\phi(\bm e_i)$ and Taylor degree $2\le r\le N$,
we estimate the degree-$r$ Taylor term and its variance proxy
using trimmed means with trimmed count $k$:
\begin{align*}
    \widehat Z_{i,r}
    &\gets
    \Trim_k(\bar Z_{i,r}^{(1)},\ldots,\bar Z_{i,r}^{(m)}),\\
    \widehat v_{i,r}
    &\gets
    \frac{\binom{2r-2}{r-1}}{(r-1)!}
    \Trim_k\bigl(
      |\bm e_i^\top\bm c_r(\bm g_1)|^2,\ldots,
      |\bm e_i^\top\bm c_r(\bm g_m)|^2
    \bigr).
\end{align*}
This gives an estimate
$\widehat s_i\gets\sum_{r=2}^{N}\sqrt{\widehat v_{i,r}}$
of the error proxy $s_i$.
We evaluate the $K$ LOO values with the largest estimated error
proxies exactly and output the sketched estimates for the rest.

\paragraph{Pseudocode.}
The pseudocode can be found in
\Cref{alg:optimal-context-attribution:loo-attribution}.

\begin{algorithm}[hbtp]
\DontPrintSemicolon
\caption{LOO Attribution}
\label{alg:optimal-context-attribution:loo-attribution}
\KwIn{The circuit $\phi$; truncation degree $N$, sketch count $m$, exact evaluations $K$, and trimmed count $k$.}
\KwOut{Leave-one-out estimates.}
Draw independent $\bm g_1,\ldots,\bm g_m\sim\mathcal{CN}(0,I_n)$.\;
\For{$j\in[m]$}{
    Compute the forward differentiation of $\grad\phi$
    at $z\bm g_j$ through degree $N-1$.\;
    Set
    \[
        \bm c_r(\bm g_j)\gets
        \frac1r[z^{r-1}]\grad\phi(z\bm g_j),
        \qquad 2\le r\le N\,.
    \]
}
\For{$i\in[n]$}{
    Compute
    \[
        \bar Z_{i,r}^{(j)}
        \gets
        \Re\frac{
        \bm e_i^\top\bm c_r(\bm g_j)
        (\bm g_j^\dagger\bm e_i)^{r-1}
        }{(r-1)!}\,,\qquad 2\le r\le N\,, j\in[m]\,.
    \]\;
    For every $2\le r\le N$, compute
    \begin{align*}
        \widehat Z_{i,r} &\gets
        \Trim_k\bigl(\bar Z_{i,r}^{(1)},\ldots,\bar Z_{i,r}^{(m)}\bigr)\,,\\
        \widehat v_{i,r} &\gets
        \frac{\binom{2r-2}{r-1}}{(r-1)!}
        \Trim_k\bigl(
          |\bm e_i^\top\bm c_r(\bm g_1)|^2,\ldots,
          |\bm e_i^\top\bm c_r(\bm g_m)|^2
        \bigr)\,.
    \end{align*}
    Set $\widehat s_i\gets\sum_{r=2}^{N}\sqrt{\widehat v_{i,r}}$.\;
}
Let $F\subseteq[n]$ index the $K$ largest values of
$\widehat s_i$, breaking ties arbitrarily.\;
Evaluate $\phi(\bm e_i)$ exactly for every $i\in F$.\;
Return estimates
\[
    \widehat L_i=
    \begin{cases}
      \phi(\bm e_i)\,, &i\in F\,,\\
      \phi(\bm0)+\bm e_i^\top\grad\phi(\bm0)
        +\sum_{r=2}^{N}\widehat Z_{i,r}\,, &i\notin F\,.
    \end{cases}
\]
\end{algorithm}

The next lemma adapts the trimming argument in the proof of
\cite[Proposition~4.5.1]{OOR25} to avoid a union bound over all $n$ LOO estimates.

\begin{lemma}[Coordinates with Small Error Proxies]
\label[lemma]{lem:optimal-context-attribution:small-proxies}
Consider \Cref{alg:optimal-context-attribution:loo-attribution}
with truncation degree $N$, sketch count $m$, and trimmed count
$1\le k< \nicefrac{m}2$.
Let $\widehat Z_{i,r}$ denote its estimate of the degree-$r$
Taylor term $T^{(r)}[\bm e_i^{\otimes r}]$, and let
$\widehat s_i$ denote its estimate of the error proxy $s_i$.
Order the coordinates so that
$s_1\ge\cdots\ge s_n$ and suppose $\sum_i s_i^2\le \nicefrac1{16}$.
For any $M>0$, with probability at least $1-4Me^{-k}$ over
the Gaussian sketches, simultaneously for every $i>M$,
\begin{align*}
    \widehat s_i&\le\frac N4\sqrt{\frac mM},\\
    \left|\sum_{r=2}^N
      \bigl(\widehat Z_{i,r}-T^{(r)}[\bm e_i^{\otimes r}]\bigr)\right|
    &\le\frac N4\sqrt{\frac mM}+\frac1{4\sqrt M}.
\end{align*}
\end{lemma}

The key idea behind \Cref{lem:optimal-context-attribution:small-proxies} is that a trimmed estimate exceeds its bound only if at least $k+1$
observations exceed the corresponding bound. We use independence
between sketches to bound this probability and then sum over
coordinates with small error proxies.

\begin{proof}
Define the unbiased estimate of the variance proxy from a
single sketch by
\[
    Y_{i,r}^{(j)}
    =\frac{\binom{2r-2}{r-1}}{(r-1)!}
      |\bm e_i^\top\bm c_r(\bm g_j)|^2.
\]
By
\Cref{item:the-toolkit:gaussian-tensor-moments-lemma:energy-moments,item:the-toolkit:gaussian-tensor-moments-lemma:sketch-moments},
\[
    \E_{\bm g_j}|\bar Z_{i,r}^{(j)}|^2\le v_{i,r},
    \qquad
    \E_{\bm g_j}Y_{i,r}^{(j)}=v_{i,r}.
\]
Fix $i>M$ and $2\le r\le N$.
If at most $k$ value observations lie outside
$[-\frac14\sqrt{m/M},\frac14\sqrt{m/M}]$,
trimming removes them. The same holds if at most $k$ estimates
$Y_{i,r}^{(j)}$ exceed $m/(16M)$.
Thus either trimmed estimate can exceed its bound only if
at least $k+1$ corresponding observations do so.

For a fixed set $J\subseteq[m]$ of $k+1$ sketches,
independence and Markov's inequality give
\begin{align*}
    \Pr_{\bm g_1,\ldots,\bm g_m}\left(
      \forall j\in J,\;
      |\bar Z_{i,r}^{(j)}|>\frac14\sqrt{\frac mM}
    \right)
    &\le\left(\frac{16M}{m}v_{i,r}\right)^{k+1},\\
    \Pr_{\bm g_1,\ldots,\bm g_m}\left(
      \forall j\in J,\;
      Y_{i,r}^{(j)}>\frac m{16M}
    \right)
    &\le\left(\frac{16M}{m}v_{i,r}\right)^{k+1}.
\end{align*}
There are $\binom m{k+1}$ possible sets $J$ and two types
of estimates.

To sum these probabilities over coordinates and degrees,
recall that decreasing order of the $s_i$'s and
$\sum_i s_i^2\le \nicefrac1{16}$ give
\[
    v_{i,r}\le s_i^2\le\frac1{16M}
    \qquad(i>M).
\]
By the definition $s_i=\sum_{r=2}^N\sqrt{v_{i,r}}$,
we also have $\sum_{r=2}^N v_{i,r}\le s_i^2$. Therefore
\begin{align}
    \sum_{i>M}\sum_{r=2}^N v_{i,r}^{k+1}
    &\le\left(\frac1{16M}\right)^k
       \sum_{i>M}\sum_{r=2}^N v_{i,r}\notag\\
    &\le\left(\frac1{16M}\right)^k\sum_{i>M}s_i^2
     \le\frac1{16^{k+1}M^k}.
    \label{eq:optimal-context-attribution:proxy-power-sum}
\end{align}

Summing over subsets of sketches, the two types of estimates,
and coordinates and degrees gives
\begin{align*}
    &\Pr_{\bm g_1,\ldots,\bm g_m}\left[
      \begin{gathered}
        \exists\,i\in[n],\ i>M,\ 2\le r\le N:\\
        |\widehat Z_{i,r}|>\frac14\sqrt{\frac mM}
        \text{ or }
        \widehat v_{i,r}>\frac m{16M}
      \end{gathered}
    \right]\\
    &\quad\le
      2\binom m{k+1}\left(\frac{16M}{m}\right)^{k+1}
      \sum_{i>M}\sum_{r=2}^N v_{i,r}^{k+1}
      \alignnote{union bound}\\
    &\quad\le
      2M\binom m{k+1}m^{-(k+1)}
      \alignnote{by \eqref{eq:optimal-context-attribution:proxy-power-sum}}\\
    &\quad\le\frac{2M}{(k+1)!}
      \alignnote{$\binom m{k+1}\le m^{k+1}/(k+1)!$}\\
    &\quad\le4Me^{-k}
      \alignnote{$(k+1)!\ge e^k/2$}.
\end{align*}

Conditioned upon all these estimates satisfying their bounds,
every $i>M$ has
\[
    \widehat s_i
    =\sum_{r=2}^N\sqrt{\widehat v_{i,r}}
    \le\frac N4\sqrt{\frac mM}.
\]
By \Cref{item:the-toolkit:gaussian-tensor-moments-lemma:sketch-moments},
\[
    |T^{(r)}[\bm e_i^{\otimes r}]|
    =|\E_{\bm g_j}\bar Z_{i,r}^{(j)}|
    \le\left(\E_{\bm g_j}|\bar Z_{i,r}^{(j)}|^2\right)^{1/2}
    \le\sqrt{v_{i,r}}.
\]
Hence
\begin{align*}
    \left|\sum_{r=2}^N
      \bigl(\widehat Z_{i,r}-T^{(r)}[\bm e_i^{\otimes r}]\bigr)\right|
    &\le\sum_{r=2}^N|\widehat Z_{i,r}|+s_i\\
    &\le\frac N4\sqrt{\frac mM}+\frac1{4\sqrt M}.
    \qedhere
\end{align*}
\end{proof}

We are now ready to prove our main result.
\begin{proof}[Proof of \Cref{thm:optimal-context-attribution}]
We first control the degree-wise estimates and their variance
proxies, keeping $N,m,K$ general.
We then use stability to bound the error after selecting $K$
queries for exact evaluation.
Finally, we choose the parameters and bound the complexity.

Suppose $\phi$ is $\alpha$-stable and $\alpha(4)<\infty$.
By scaling $\phi$, we may assume without loss of generality
that $\alpha(4)=1$.

\smallskip
\noindent\emph{Degree-Wise Estimates.}
Apply \Cref{item:the-toolkit:gaussian-tensor-moments-lemma:sketch-moments}
to the symmetric order-$(r-1)$ tensor
$T^{(r)}[\bm e_i,\cdot]$ and the direction $\bm e_i$.
The degree-$r$ estimates are unbiased and satisfy
\[
    \E_{\bm g_j}\bar Z_{i,r}^{(j)}
    =T^{(r)}[\bm e_i^{\otimes r}],
    \qquad
    \Var_{\bm g_j}(\bar Z_{i,r}^{(j)})\le v_{i,r}.
\]
For the scaled squared contractions $Y_{i,r}^{(j)}$ defined
in the proof of
\Cref{lem:optimal-context-attribution:small-proxies},
\Cref{item:the-toolkit:gaussian-tensor-moments-lemma:energy-moments}
gives
\begin{align*}
    \E_{\bm g_j}Y_{i,r}^{(j)}
    &=\binom{2r-2}{r-1}
      \norm{T^{(r)}[\bm e_i,\cdot]}_F^2=v_{i,r},\\
    \Var_{\bm g_j}(Y_{i,r}^{(j)})
    &\le\binom{2r-2}{r-1}
      \left(\E_{\bm g_j}Y_{i,r}^{(j)}\right)^2
      =\binom{2r-2}{r-1}v_{i,r}^2.
\end{align*}
Thus the same sketches estimate both the degree-$r$ value
and its variance proxy.

\smallskip
\noindent\emph{Total Variance.}
By stability at $B=4$,
$\norm{T^{(r)}}_F\le\alpha(4)/4^r$.

Since
$\sum_i\norm{T^{(r)}[\bm e_i,\cdot]}_F^2=\norm{T^{(r)}}_F^2$,
Minkowski's inequality gives
\begin{align}
    \left(\sum_{i=1}^n s_i^2\right)^{1/2}
    &\le
    \sum_{r=2}^{N}
      \sqrt{\binom{2r-2}{r-1}}\norm{T^{(r)}}_F
    \le
    \alpha(4)\sum_{r=2}^{\infty}\frac{2^{r-1}}{4^r}
    =\frac{\alpha(4)}4.
    \label{eq:optimal-context-attribution:total-proxy-bound}
\end{align}

Thus $\sum_i s_i^2\le1/16$.
For the analysis, order the coordinates by decreasing $s_i$ so that $s_i \geq s_{i+1}$ for all $1\leq i<n$.

\smallskip
\noindent\emph{Failure Probability.}
We will apply
\Cref{prop:context-attribution:trimmed-mean-concentration}
to estimate the Taylor terms and variance proxies among the
first $M$ coordinates, and
\Cref{lem:optimal-context-attribution:small-proxies}
to control the remaining coordinates.
The algorithm uses truncation degree $N$, sketch count $m$,
and $K$ exact evaluations. Choose
\begin{equation}
\label{eq:optimal-context-attribution:parameters}
\begin{gathered}
    N=\log_4(\nicefrac1\eps),\qquad
    m=\frac{256}{\eps\sqrt N}
       \left(N+\log(\nicefrac1\delta)\right),\qquad
    K=\frac{16\sqrt N}{\eps},\\
    M=\frac{N^2m}{\eps^2},
    \qquad
    k=\log(\nicefrac{64MN}\delta).
\end{gathered}
\end{equation}
We assume $N\ge2$, as we compute gradients exactly.
The trimmed-mean application requires
\begin{equation}
\label{eq:optimal-context-attribution:sample-size}
    m\ge64\log(\nicefrac{64MN}\delta),
    \qquad
    m\ge64\binom{2N-2}{N-1}\log(\nicefrac{64MN}\delta).
\end{equation}
We defer checking these conditions and first bound the failure
probabilities and consider the conclusions conditioned upon success.

Assuming these conditions, apply the trimmed-mean guarantee
with failure probability $\delta/(4MN)$ to each value and
variance-proxy estimate among the first $M$ coordinates.
A union bound over at most $2MN$ estimates gives total failure
probability at most
\[
    2MN\cdot\frac{\delta}{4MN}=\frac\delta2.
\]
For the remaining coordinates,
\Cref{lem:optimal-context-attribution:small-proxies}
gives failure probability at most
\begin{align*}
    4Me^{-k}
    =\frac{\delta}{16N}
    \le\frac\delta2\,.
    \alignnote{choice of $k=\log(64MN/\delta)$ \eqref{eq:optimal-context-attribution:parameters}}
\end{align*}
Thus, subject to the sample-size conditions, both conclusions
hold simultaneously with probability at least $1-\delta$.
We now work conditioned upon success of both applications.

\smallskip\noindent\emph{Estimation Error.}
For every $i\le M$ and $2\le r\le N$,
\Cref{prop:context-attribution:trimmed-mean-concentration}
gives
\begin{align*}
    |\widehat Z_{i,r}-T^{(r)}[\bm e_i^{\otimes r}]|
    &\le7\sqrt{\frac{v_{i,r}\log(\nicefrac{64MN}\delta)}m}\,,\\
    |\widehat v_{i,r}-v_{i,r}|
    &\le7v_{i,r}\sqrt{
      \frac{\binom{2r-2}{r-1}\log(\nicefrac{64MN}\delta)}m}
    \le\frac78v_{i,r}
      \alignnote{$m\ge64\binom{2N-2}{N-1}\log(64MN/\delta)$ \eqref{eq:optimal-context-attribution:sample-size}}\,.
\end{align*}
Thus $v_{i,r}/8\le\widehat v_{i,r}\le2v_{i,r}$, and
\begin{align}
    \widehat s_i
    &=\sum_{r=2}^N\sqrt{\widehat v_{i,r}}
      \ge\frac{s_i}{\sqrt8}\,,\notag\\
    \sum_{i\le M}\widehat s_i^2
    &\le2\sum_{i\le M}s_i^2
      \le\frac18
      \alignnote{total variance bound \eqref{eq:optimal-context-attribution:total-proxy-bound}}\,.
      \label{eq:optimal-context-attribution:squared-proxy-estimate}
\end{align}

For every $i>M$,
\Cref{lem:optimal-context-attribution:small-proxies}
gives
\begin{align*}
    \widehat s_i
    &\le\frac N4\sqrt{\frac mM}
    =\frac\eps4
      \alignnote{choice of $M=N^2m/\eps^2$ \eqref{eq:optimal-context-attribution:parameters}}\,,\\
    \left|\sum_{r=2}^N
      \bigl(\widehat Z_{i,r}-T^{(r)}[\bm e_i^{\otimes r}]\bigr)\right|
    &\le\frac N4\sqrt{\frac mM}+\frac1{4\sqrt M}
    = \frac\eps4+\frac{\eps}{4N\sqrt m}
      \alignnote{choice of $M=N^2m/\eps^2$ \eqref{eq:optimal-context-attribution:parameters}}\\
    &\le\frac\eps2
      \alignnote{$N\sqrt m\ge1$}\,.
\end{align*}

We now argue that the algorithm evaluates every coordinate
satisfying
\[
    \widehat s_i>
    \max\left\{\frac\eps4,\frac1{\sqrt{8K}}\right\}
\]
exactly.
Every such coordinate belongs to the first $M$, since
$\widehat s_i\le\eps/4$ for $i>M$.
Hence
\begin{align*}
    \left|\left\{i\le M:
      \widehat s_i>\frac1{\sqrt{8K}}\right\}\right|
    \le8K\sum_{i\le M}\widehat s_i^2
    \le K.
    \alignnote{$\sum_{i\le M}\widehat s_i^2\le\nicefrac18$ \eqref{eq:optimal-context-attribution:squared-proxy-estimate}}
\end{align*}
Thus at most $K$ coordinates exceed the displayed cutoff.
The algorithm selects the $K$ largest estimated proxies,
so it evaluates all these coordinates exactly.
Consequently, every coordinate left unevaluated satisfies
\[
    \widehat s_i\le
    \max\left\{\frac\eps4,\frac1{\sqrt{8K}}\right\}.
\]
For an unevaluated $i\le M$, substituting
$s_i\le\sqrt8\,\widehat s_i$ into the degree-wise error bound
then gives
\begin{align*}
    \left|\sum_{r=2}^N
      \bigl(\widehat Z_{i,r}-T^{(r)}[\bm e_i^{\otimes r}]\bigr)\right|
    &\le7s_i\sqrt{\frac{\log(\nicefrac{64MN}\delta)}m}\\
    &\le\max\left\{
        7\eps\sqrt{\frac{\log(\nicefrac{64MN}\delta)}{2m}},
        7\sqrt{\frac{\log(\nicefrac{64MN}\delta)}{mK}}
      \right\}.
\end{align*}
We will verify below that
\[
    m\ge128\log(\nicefrac{64MN}\delta),
    \qquad
    mK\ge\frac{196}{\eps^2}\log(\nicefrac{64MN}\delta).
\]
This ensures that the two terms are at most $\nicefrac{7\eps}{16}$ and $\nicefrac\eps2$,
respectively. Thus every unevaluated coordinate has sampling
error at most $\nicefrac\eps2$.
Adding the Taylor remainder $\nicefrac\eps3$ from
\Cref{lem:the-toolkit:taylor-truncation} gives total error at
most $\eps$.

\smallskip\noindent\emph{Evaluation Count.}
In order to apply
\Cref{prop:context-attribution:trimmed-mean-concentration}
to bound the degree-wise value and variance-proxy estimates,
we require
\[
    m\ge64\log(\nicefrac{64MN}\delta),
    \qquad
    m\ge64\binom{2N-2}{N-1}\log(\nicefrac{64MN}\delta).
\]
The second condition makes the relative error of the
variance-proxy estimates at most $7/8$.

We first bound the logarithmic term.
\begin{align}
    \log\left( \frac{64MN}\delta \right)
    &=\log\left(\frac{64N^3m}{\eps^2\delta}\right)
      \alignnote{choice of $M=N^2m/\eps^2$
        \eqref{eq:optimal-context-attribution:parameters}}\notag\\
    &\le\log\left(
      \frac{2^{14}N^{7/2}}{\eps^3\delta^2}
      \right)
      \alignnote{choice of $m\le256\sqrt N/(\eps\delta)$
        \eqref{eq:optimal-context-attribution:parameters}}\notag\\
    &\le14\log2+4\log N
      +3\log(1/\eps)+2\log(1/\delta)
      \alignnote{$\nicefrac72\le4,\ N\ge2$}\notag\\
    &\le18\log N+6N\log2+2\log(1/\delta)
      \alignnote{choice of $N=\log_4(\nicefrac1\eps)$
        \eqref{eq:optimal-context-attribution:parameters},
        $N\ge2$}\notag\\
    &\le11\left(N+\log(1/\delta)\right)
      \alignnote{$\log N\le N/e$}\,.
\end{align}
The central-binomial bound gives
\[
    \binom{2N-2}{N-1}
    \le\frac{4^{N-1}}{\sqrt N}
    =\frac1{4\eps\sqrt N}.
\]
Therefore,
\begin{align}
    64\binom{2N-2}{N-1}\log\left( \frac{64MN}\delta \right)
    \leq \frac{16}{\eps\sqrt{N}}\cdot \log\left( \frac{64MN}\delta \right)
    \le\frac{176}{\eps\sqrt N}
       \left(N+\log(1/\delta)\right)
    \le m\,.
    \label{eq:optimal-context-attribution:sketch-bound}
\end{align}
Since $\binom{2N-2}{N-1}\ge2$ for $N\ge2$, this also proves
$m\ge128\log(64MN/\delta)$.
In particular, both sample-size conditions hold.

We next check the product $mK$ used in the sampling-error bound.
Our choices give
\begin{align*}
    mK
    &\ge\frac{16}{\eps\sqrt N}
      \log(\nicefrac{64MN}\delta)\,K
      \alignnote{bound on $m$
        \eqref{eq:optimal-context-attribution:sketch-bound}}\\
    &\ge\frac{256}{\eps^2}\log(\nicefrac{64MN}\delta)
      \alignnote{choice of $K=16\sqrt{N}/\eps$
        \eqref{eq:optimal-context-attribution:parameters}}.
\end{align*}
This verifies the second bound used to control the maximum
of the two sampling-error terms and concludes the proof of \Cref{item:optimal-context-attribution:accuracy}.

\smallskip
\noindent\emph{Space, Time, \& Query Complexity.}
By \Cref{prop:the-toolkit:backward-differentiation,prop:the-toolkit:forward-differentiation},
each backward--forward sketch computes all retained coefficients
in $O(\size(\phi)N\log N)$ arithmetic operations and
$O(\size(\phi)N)$ auxiliary space.
Thus the $m$ sketches take $O(\size(\phi)mN\log N)$
arithmetic operations.
Using successive powers, forming all observations takes $O(nmN)$
operations. Each trimmed mean can be computed in $O(m)$ time
by selection, and the exact evaluations take $O(\size(\phi)K)$
operations.
Since $\size(\phi)\ge n$, the total is
\[
    O\bigl(\size(\phi)(mN\log N+K)\bigr).
\]
Each sketch stores one length-$n$ direction and $N-1$
length-$n$ coefficient vectors, using $O(nmN)$ space in total.
Together with the Taylor evaluations, this uses auxiliary space
\[
    O\bigl(\size(\phi)N+nmN\bigr).
\]

Substituting the parameters proves
\Cref{item:optimal-context-attribution:complexity}.
A single sketch returns every retained degree simultaneously,
so estimating the degrees separately does not increase the
number of model calls.
The total number of evaluation calls is
\[
    \min\{n,m+K\}
    =O\left(\min\left\{
      n,\frac{\sqrt N+\log(\nicefrac1\delta)/\sqrt N}{\eps}
    \right\}\right).
\]
Since $N=\Theta(\log(\nicefrac1\eps))$, this is
\[
    O\left(\min\left\{
      n,\frac{\sqrt{\log(\nicefrac1\eps)}+\log(\nicefrac1\delta)}{\eps}
    \right\}\right),
\]
which proves \Cref{item:optimal-context-attribution:evaluations}.
\end{proof}

\subsection{Query Complexity Lower Bound}
\label{sec:optimal-context-attribution:lower-bound}

To match the evaluation count in
\Cref{item:optimal-context-attribution:evaluations}, we prove a lower
bound for estimating all diagonal entries simultaneously in a black-box oracle model.

\begin{definition}[Oracle Access]
\label[definition]{def:optimal-context-attribution:oracle-access}
We assume we can obtain $\phi(\bm0)$ and $\grad\phi(\bm0)$ for free.
For $\bm w,\bm g\in\C^n$ and an integer $N\ge1$, the oracle provides
\[
    (\bm w,\bm g,N)\longmapsto
    \bigl(
      \phi(\bm w),\grad\phi(\bm w),
      D_{\bm g}^1\grad\phi(\bm w),\ldots,
      D_{\bm g}^N\grad\phi(\bm w)
    \bigr).
\]
Each query counts as one unit, regardless of $N$, and its inputs
may depend on all previous responses.
\end{definition}
These queries include the backward--forward sketches and exact
evaluations used by our algorithm.
We emphasize that this is a black-box oracle lower bound and does not
rule out algorithms with unrestricted access to the circuits.

\begin{restatable}[Simultaneous Diagonal Estimation]{lemma}{simultaneousdiagonallowerbound}
\label[lemma]{lem:optimal-context-attribution:diagonal-lower-bound}
For every $d\ge64$, consider any randomized adaptive procedure
that estimates the diagonal of a real symmetric $d\times d$
matrix using at most $\nicefrac{d}8$ real matrix--vector queries.
There is a matrix $A$ with $\norm{A}_F\le\nicefrac1{16}$ for which
\[
    \Pr_R\left(
      \max_{i\in[d]}|\widehat L_i-A_{ii}|>
      \frac{\sqrt{\log d}}{128d}
    \right)\ge\frac12,
\]
where $R$ is the procedure's internal randomness.
\end{restatable}

\emph{Proof sketch.}
As in the proof of \citet[Lemma~3{;} arXiv version]{amsel2024fixed}, we condition
on the queries and their responses and use the remaining randomness
to bound the estimation error. We use a symmetric Gaussian matrix
to bound the maximum error over the diagonal entries.
Take $A=G/(64d)$, where $G$ is a symmetric Gaussian matrix
whose entries on and above the diagonal are independent, with
$G_{ii}\sim\mathcal N(0,2)$ and $G_{ij}\sim\mathcal N(0,1)$
for $i<j$. For simplicity, suppose the queries are the first $m$ standard basis vectors
$\bm e_1,\ldots,\bm e_m$, where $m\le d/8$.
Each response $A\bm e_j$ reveals the $j$-th column.
Conditioned on these responses and the internal randomness $R$,
all the estimates $(\widehat L_i\mid R, A\bm e_1, \dots, A\bm e_m)$ for $i\in [d]$ are deterministic.
On the other hand,
conditioned on these columns $A\bm e_1, \dots, A\bm e_m$,
the remaining diagonal entries satisfy
\[
    A_{ii}\overset{\mathrm{iid}}{\sim}
    \mathcal N\left(0,\frac{2}{(64d)^2}\right),
    \qquad m<i\le d.
\]
For $m<i\leq d$,
the best possible deterministic estimate that maximizes the probability of the accuracy event is $\widehat L_i = 0$.
The largest absolute error among the $d-m$ entries is essentially the same as the maximum of $\Omega(d)$ independent Gaussian variables.
Hence even in the best case,
the maximum error is of order $\sqrt{\log d}/d$ with constant probability.

For general queries, we can rotate the basis so that the
queried vectors span the first few coordinate directions.
A Gaussian conditioning argument due to
\citet[Lemma~3.4 and Appendix~C.2]{SER18} shows that the remaining
square block is still an independent symmetric Gaussian matrix, even
for adaptive queries. 
\citet[Lemma~C.1 and Theorem~4.2]{JPWZ21} use this block to bound the
randomness in the trace.
Here the diagonal depends on the basis.
After rotating back, we show that at least $\nicefrac{d}2$ diagonal
entries in the original basis each contain an independent
Gaussian component of standard deviation $1/(64d)$. Even if the
procedure knows the conditional means of these entries, this noise
gives the same lower bound as in the standard basis example.
We give the formal proof in \Cref{sec:simultaneous-diagonal-estimation-proof}.

\begin{theorem}[Oracle Query Lower Bound]
\label{thm:optimal-context-attribution:oracle-lower-bound}
For all sufficiently small $\eps>0$ and $n\ge1/\eps$, any randomized
adaptive procedure that, for every real symmetric $n\times n$
matrix $A$ with $\norm{A}_F\le \nicefrac1{16}$, estimates the LOO values of
$\phi_A(\bm w)=\bm w^\top A\bm w$ with
\[
    \Pr_R\left(
      \max_{i\in[n]}|\widehat L_i-A_{ii}|\le\eps
    \right)\ge\frac23
\]
requires
\[
    \Omega\left(\min\left\{
      n,\frac{\sqrt{\log(\nicefrac1\eps)}}{\eps}
    \right\}\right)
\]
queries in the oracle model of
\Cref{def:optimal-context-attribution:oracle-access}.
Here the probability is over the procedure's internal randomness $R$.
These functions $\phi_A$ are $\alpha$-stable for
$\alpha(B)=B^2/16$, so $\alpha(4)=1$.
\end{theorem}

\begin{proof}
For a quadratic $\phi_A(\bm w)=\bm w^\top A\bm w$,
\[
    \grad\phi_A(\bm w)=2A\bm w,\qquad
    D_{\bm g}^1\grad\phi_A(\bm w)=2A\bm g,\qquad
    D_{\bm g}^r\grad\phi_A(\bm w)=0\quad(r>1).
\]
The oracle response is therefore determined by $A\bm w$ and
$A\bm g$.
Note that each complex matrix--vector product can be simulated using two
real products, by applying $A$ separately to the real and
imaginary parts.
Thus one oracle query can be simulated using a constant number of real
matrix--vector queries.

Let $q$ be the worst-case number of oracle queries.
If $q>n/64$, the claimed bound follows from $q=\Omega(n)$.

Otherwise, take $d=64q\le n$. Given a real symmetric $d\times d$
matrix in the Frobenius ball, pad it with zeros and run the
procedure in dimension $n$. The first $d$ LOO values are the
diagonal entries of the original matrix. By the simulation above,
the procedure uses at most $4q=\nicefrac{d}{16}\le \nicefrac{d}8$ real matrix--vector
queries to estimate them. Thus
\Cref{lem:optimal-context-attribution:diagonal-lower-bound}
gives a matrix on which the maximum error exceeds
$\sqrt{\log d}/(128d)$ with probability at least $\nicefrac{1}2$.
The procedure succeeds to error $\eps$ with probability at least
$\nicefrac{2}3$ on every input, so
\[
    \eps>\frac{\sqrt{\log d}}{128d}.
\]

Since $\sqrt{\log d}\ge1$, this gives $\nicefrac1\eps<128d$ and hence
$\log(\nicefrac1\eps)<\log(128d)\le3\log d$. Substituting $d=64q$
into the error bound and rearranging gives
\[
    q>\frac{\sqrt{\log d}}{8192\eps}
      \ge\frac{\sqrt{\log(1/\eps)}}{8192\sqrt3\,\eps}.
\]
Together the two cases give the claimed lower bound.

Finally, $T_{\phi_A}^{(2)}=A$ and every other Taylor tensor is zero.
Since $\norm{A}_F\le\nicefrac1{16}$, these functions are $\alpha$-stable
for $\alpha(B)=\nicefrac{B^2}{16}$, with $\alpha(4)=1$.
\end{proof}

For constant failure probability, this matches the evaluation count
in \Cref{item:optimal-context-attribution:evaluations}.

\section{Experiments}
\label{sec:experiments}
In this section, we examine Taylor coefficient decay and compare
backward--forward sketches with forward-only sketches on a small
retrieval task.
We also evaluate how well the standard deviation proxy tracks
the empirical standard deviation across trials.

\paragraph{Setup.}
We perform LOO attribution for entire context chunks rather than
individual tokens.
The only adjustment from \Cref{sec:optimal-context-attribution} is that we
assign one downweight $w_i$ to all tokens in chunk~$i$.
We prompt \texttt{Llama-3.2-1B-Instruct}\footnote{\url{https://huggingface.co/unsloth/Llama-3.2-1B-Instruct}} at 64-bit precision with a retrieval task whose answer is in one of 12 chunks (the 7th chunk) and all other chunks contain irrelevant information.
We show the full prompt below.

\begin{lstlisting}[basicstyle=\small\ttfamily, columns=fullflexible, breaklines=true, frame=single]
System: Answer the question using only the provided text.
User: Documents:
[1] The Calder Museum has opened an exhibition of ancient coins recovered
    from a shipwreck off the Pellan coast. Divers spent two summers bringing
    the pieces to the surface, and conservators cleaned each one by hand.
    The exhibition will stay open until the end of the year, and entry is
    free for children under twelve.
[2] The most borrowed book at the Pinecrest public library this year was a
    novel about a young painter who leaves home to find her lost brother.
    The librarians say they had to order six extra copies to keep up with
    demand. The author, Celia Moraine, will visit the library next season to
    read from the book.
[3] Four climbers from the Ardel Alpine Club reached the summit of Mount
    Keldar after a long ascent of its northern ridge. Their food and tents
    were carried up the lower slopes by a string of camels hired from a
    nearby village. The expedition leader, Sigrun Halvo, said the view from
    the top was worth every freezing night on the ice.
[4] From next month, the overnight train between Ostbury and Lanmere will
    run only on Fridays. The railway company says the change will free up
    carriages for its busy coastal routes. Passengers who already hold
    tickets for other nights can exchange them at any station without paying
    a fee. A new timetable will be posted on the station noticeboards soon.
[5] On the first morning of their trip down the Ormsby River, Marta Jensen
    caught a trout that weighed almost four kilograms. Her brother had
    brought a new rod but spent most of the day untangling his line. They
    cooked their catch over a small fire and returned to the cabin before
    dark.
[6] At the Eastfield garden show, first prize went to a giant turnip grown
    by Walter Pemberton, who has kept an allotment for forty years. He said
    the secret was rainwater collected in barrels and a lot of patience. The
    judges also praised his neat rows of herbs and the hand-painted signs
    along his paths.
[7] Chemists at the Varnell Institute have created a ceramic coating that
    turns blue when it is heated above four hundred degrees. The coating
    could warn engineers when pipes in power plants begin to overheat.
    Professor Tomasz Ellery, who led the project, said the change is fully
    reversible and survives thousands of heating cycles.
[8] Last spring the Hollowbrook bakery introduced a new pastry filled with
    lemon jam and toasted almonds. The head baker, Imelda Corvan, spent
    three months testing the recipe in a small kitchen behind the shop. On
    opening day the queue stretched past the post office, and the first tray
    sold out in under an hour. Regular customers now reserve a box every
    weekend.
[9] The lantern festival in Sorrel Bay is held every year in April, when the
    evenings are mild and the bay is calm. Thousands of paper lanterns are
    released over the water at sunset, and the streets are decorated with
    ribbons. Visitors are advised to book rooms early because the town has
    only three small hotels.
[10] A new statue of the poet Edwin Larke now stands in Harlow Square. The
     sculptor cast it in bronze at a small foundry outside the city, and it
     took nine months to finish. On the unveiling day the mayor read one of
     Larke's poems aloud, and pigeons settled on the statue's shoulders
     within minutes.
[11] The spring concert at Marlowe Hall featured a young soloist, Yuki
     Hartmann, playing a rarely heard piece for oboe and orchestra. The
     audience gave her a standing ovation that lasted nearly five minutes.
     Critics praised her calm stage presence and the clarity of the slow
     movement.
[12] The winter soup at the Dovecote Inn is made with barley, smoked ham and
     plenty of black pepper. The recipe has been handed down by the
     innkeeper's family for four generations. Travellers who arrive cold from
     the moors often order a second bowl, and the inn sells jars of it at the
     counter.
Question: What color does the Varnell Institute coating turn when it is
heated?
\end{lstlisting}

The evaluation function $\phi$ is the sum of the token
log-probabilities of the following completion, generated once
with greedy decoding. We hold this completion fixed across
deletion weights and trials.

\begin{lstlisting}[basicstyle=\small\ttfamily, columns=fullflexible, breaklines=true, frame=single]
Assistant: The Varnell Institute coating turns blue when it is heated above
400 degrees
\end{lstlisting}

We implement backward--forward sketches in JAX, using its
\href{https://docs.jax.dev/en/latest/jax.experimental.jet.html}
{\texttt{jet}} module to apply forward differentiation to the
value-and-gradient computation of $\phi$.
We use \texttt{complex128} for complex arithmetic.
Each trial runs on one NVIDIA H100 GPU.

Empirically, context attributions are often concentrated on a few
sources~\citep{cohenwang2024contextcite,liu2025attribot}.
Prior work uses gradient norms and embedding similarities
to rank influential sources~\citep{liu2025attribot}.
In our setting, chunk~7 contains the answer, so deleting it is
expected to strongly affect the completion's log-probability.
We perform a standard pre-processing step by ranking queries by $|\partial_i\phi(\bm0)|$ and evaluate the
query with the largest score, chunk~7, exactly.
We then set $g_7=0$ and sketch the remaining indices.
The other coordinates are independent standard complex Gaussians.
This applies the sketch to the restricted function
$\phi\big|_{w_7=0}$, with chunk~7 kept in the prompt.
The stability assumption in \Cref{item:optimal-context-attribution:accuracy}
applies to this restricted function.

\begin{figure}[htbp]
  \centering
  \newcommand{\panel}[2]{%
    \begin{subfigure}[t]{0.5\textwidth}\centering
      \includegraphics[width=\linewidth,height=0.3\textheight,keepaspectratio]{fig/#1}
      \caption{#2}
    \end{subfigure}}
  \panel{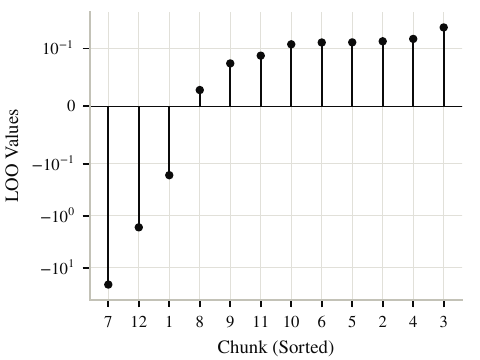}{Leave-one-out values}\hfill
  \panel{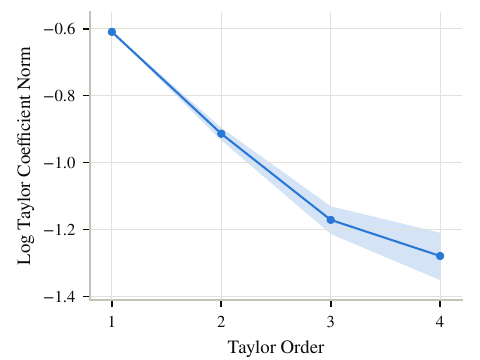}{Taylor coefficient norms}
  \caption{\textbf{The restricted function exhibits coefficient decay
  consistent with stability.}
  (a) Exact LOO differences $\phi(\bm e_i)-\phi(\bm0)$, sorted by value.
  (b) Estimated log Frobenius norms of the Taylor tensors of
  $\phi\big|_{w_7=0}$, for $r=1,\ldots,4$.
  Shading shows a 90\% bootstrap interval.}
  \label{fig:experiments:leave-one-out-values}
\end{figure}

\begin{figure}[htbp]
  \centering
  \newcommand{\panel}[2]{%
    \begin{subfigure}[t]{0.333\textwidth}\centering
      \includegraphics[width=\linewidth,height=0.18\textheight,keepaspectratio]{fig/#1}
      \caption{#2}
    \end{subfigure}}
  \newcommand{\chunkrow}[2]{%
    \panel{loo_taylor_chunk#1}{Chunk #2, Taylor}\hfill
    \panel{loo_sketch_chunk#1}{Chunk #2, backward sketch}\hfill
    \panel{loo_gunn_chunk#1}{Chunk #2, forward sketch}}
  \chunkrow{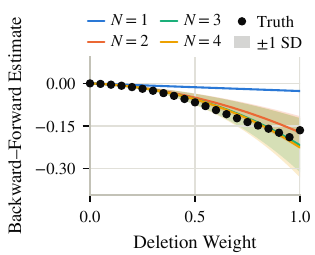}{1}\\[4pt]
  \chunkrow{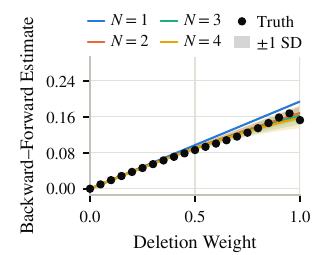}{4}\\[4pt]
  \chunkrow{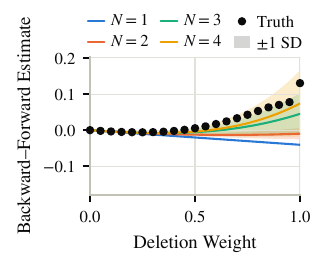}{6}\\[4pt]
  \chunkrow{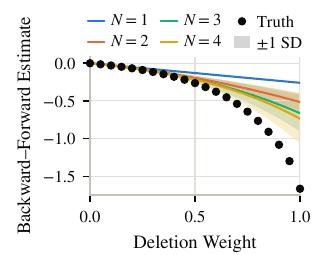}{12}
  \captionsetup{skip=4pt}
  \caption{\textbf{Backward--forward sketches have lower variance compared to forward-only sketches
  at the same sketch budget.}
  Rows correspond to chunks $1,4,6,12$.
  Columns show reference Taylor polynomials, backward--forward
  estimates, and forward-sketch estimates for $N=1,\ldots,4$.
  Dots denote direct evaluations of
  $\phi(z\bm e_i)-\phi(\bm0)$.
  Sketch curves show trial means with one standard deviation shaded.}
  \label{fig:experiments:leave-one-out-taylor-sketch}
\end{figure}

\begin{figure}[htbp]
  \centering
  \newcommand{\panel}[2]{%
    \begin{subfigure}[t]{0.5\textwidth}\centering
      \includegraphics[width=\linewidth,height=0.3\textheight,keepaspectratio]{fig/#1}
      \caption{#2}
    \end{subfigure}}
  \panel{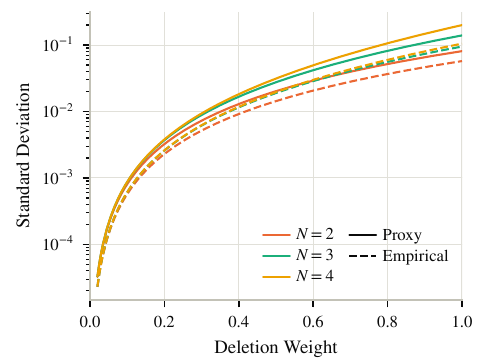}{Chunk 1}\hfill
  \panel{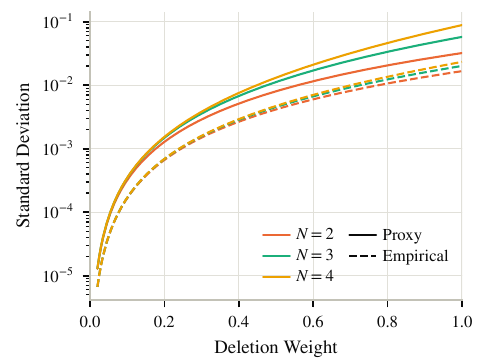}{Chunk 4}\\[4pt]
  \panel{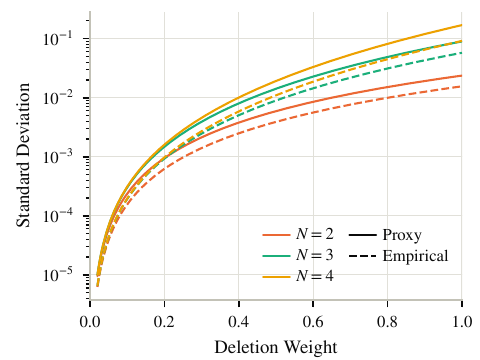}{Chunk 6}\hfill
  \panel{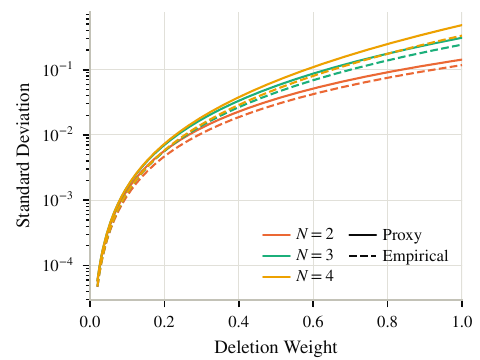}{Chunk 12}
  \caption{\textbf{The standard deviation proxy closely tracks
  the empirical standard deviation.}
  Solid curves show the mean proxy across trials; dashed curves
  show the empirical standard deviation across trials.
  Panels correspond to chunks $1,4,6,12$, with Taylor degrees
  $N=2,3,4$.}
  \label{fig:experiments:leave-one-out-standard-deviation-proxies}
\end{figure}

We run 128 independent trials with $m=8$ sketches per trial.
For each chunk and degree, each trimmed mean removes the
smallest and largest observations ($k=1$) and averages the
remaining six.
We report Taylor degrees $N=1,\ldots,4$.
\Cref{fig:experiments:leave-one-out-values,fig:experiments:leave-one-out-taylor-sketch,fig:experiments:leave-one-out-standard-deviation-proxies} show the leave-one-out values and restricted Taylor coefficient norms, the Taylor approximations against both sketches, and the standard deviation proxies, respectively.
The experiments in this section are extremely
minimal, and intended as a proof of concept.
For this small setting of $n=12$,
it may be faster to just evaluate all LOO values exactly with brute force.
For fixed accuracy and failure probability, the number of sketches
is independent of $n$, while direct evaluation takes $n$ model calls.

\paragraph{Exact Values \& Stability.}
\Cref{fig:experiments:leave-one-out-values} shows that removing
chunk 7 causes the largest decrease in the
completion's log-probability.
Panel~(a) plots the exact differences
$\phi(\bm e_i)-\phi(\bm0)$ for all twelve chunks, sorted by value
on a symmetric logarithmic scale.
Panel~(b) shows the estimated log Frobenius norms of the Taylor
tensors of the restricted function $\phi\big|_{w_7=0}$.
We estimate these norms from all 1,024 Gaussian directions
using the second-moment identity in
\Cref{item:the-toolkit:gaussian-tensor-moments-lemma:energy-moments}.
The shaded region is a 90\% bootstrap interval computed from
1,000 resamples.
The estimated norms decrease over the four displayed orders.

\paragraph{Estimates via Sketching.}
\Cref{fig:experiments:leave-one-out-taylor-sketch} compares
reference Taylor polynomials with our backward--forward estimates
and the forward-sketch baseline
(\Cref{thm:context-attribution:forward-sketch-baseline}).
For a deletion weight $z\in[0,1]$, the target is
$\phi(z\bm e_i)-\phi(\bm0)$, and its degree-$N$ Taylor
approximation is
\[
    \sum_{r=1}^{N}z^rT^{(r)}[\bm e_i^{\otimes r}].
\]
We compute the reference coefficients by numerical contour
integration, checking agreement across integration radii.
The sketch curves show the mean across trials before exact
correction, with one empirical standard deviation shaded.
At the same sketch budget, backward--forward sketches have smaller
standard deviations compared to forward-only sketches.

\paragraph{Error Proxy.}
\Cref{fig:experiments:leave-one-out-standard-deviation-proxies}
compares the empirical standard deviation across trials with
the estimated standard deviation proxy
\[
    \sum_{r=2}^{N}z^r\sqrt{\frac{\widehat v_{i,r}}m}.
\]
The factor $z^r$ accounts for the deletion weight in the
degree-$r$ term.
We average this proxy across trials and compare it with the
standard deviation of the corresponding sketch estimates.
For the displayed chunks and degrees, the proxy stays above
the empirical standard deviation and tracks its growth as
$z$ increases.
The proxy describes sampling variability.
Taylor truncation and bias from trimming are separate sources
of prediction error.

\clearpage
\section*{Acknowledgements}
We thank Sam Gunn for feedback on an initial draft that helped
improve the presentation.

\section*{AI Methodology}
We used Gemini 4 and ChatGPT 5.6 Pro / Astra to brainstorm
technical approaches, carry out calculations, and help write
code for the experiments. The authors have verified the correctness
of the results and take responsibility for them.

\printbibliography

\clearpage
\appendix
\crefalias{section}{appendix}
\crefalias{subsection}{appendix}

\section{Forward Sketch Baseline}

As a baseline for \Cref{thm:optimal-context-attribution},
we state an adaptation of \citeauthor{gunn2026sketch}'s data attribution algorithm via forward differentiation sketching to our context attribution setting.
\begin{theorem}[Forward-Sketch Baseline;
Adapted From {\cite[Theorem~2]{gunn2026sketch}}]
\label{thm:context-attribution:forward-sketch-baseline}
Let $\phi:\C^n\to\C$ be an $\alpha$-stable arithmetic circuit
that is real-valued on $[0,1]^n$, with $\alpha(4)<\infty$.
For any $\eps,\delta\in(0,1)$, forward sketches give estimates
satisfying
\[
    \max_{i\in[n]}|\widehat L_i-\phi(\bm e_i)|
    \le\eps\alpha(4)
\]
with probability at least $1-\delta$ over the sketch randomness,
using $\Ot(\size(\phi)\log(\nicefrac n\delta)/\eps^2)$ operations and $\Ot(\size(\phi) + n\log(\nicefrac{n}\delta)/\eps^2)$ auxiliary space.
\end{theorem}

\section{Continuous Masking}
\label[appendix]{sec:continuous-masking}
We implement the downweights $\bm w\in[0,1]^n$ of
\Cref{sec:optimal-context-attribution} as a continuous attention mask.
Fix the model parameters and the tokenized input of length $L$,
and let $p_1<\cdots<p_n$ be the positions of the context tokens
$\tau_1,\ldots,\tau_n$.
For each position $j\in[L]$, define its survival weight
\[
    \chi_j(\bm w)\coloneqq
    \begin{cases}
        1-w_i, & \text{if $j=p_i$},\\
        1,     & \text{otherwise},
    \end{cases}
\]
so $w_i=0$ keeps token $\tau_i$ and $w_i=1$ masks it.

Every attention head in every layer uses these weights.
At query position $t$, a head outputs
\[
    \bm o_t(\bm w)\coloneqq
    \frac{\sum_{j=1}^L M_{tj}\,\chi_j(\bm w)\exp(a_{tj}(\bm w))\,\bm u_j(\bm w)}
         {\sum_{j=1}^L M_{tj}\,\chi_j(\bm w)\exp(a_{tj}(\bm w))},
\]
where $a_{tj}(\bm w)$ and $\bm u_j(\bm w)$ are the query--key logits
and value vectors, and $M_{tj}\in\{0,1\}$ is the fixed causal mask.
We assume every attention row attends to at least one position
outside the context, such as a beginning-of-sequence token, so the
denominator is positive on $[0,1]^n$.
If the logits and values are holomorphic in $\bm w$, then so is
$\bm o_t$ wherever the denominator is nonzero.

Setting $\bm w=\bm e_i$ removes the terms for token $\tau_i$ from
both sums, giving ordinary softmax attention over the remaining tokens.
Since all other operations act on each position separately, the
remaining hidden states agree layer by layer with a run on the prompt
with $\tau_i$ deleted, keeping the positions of the other tokens fixed.
Hence $\phi(\bm e_i)$ is exactly the LOO value of token $\tau_i$.

\section{Simultaneous Diagonal Estimation}
\label{sec:simultaneous-diagonal-estimation-proof}

\simultaneousdiagonallowerbound*

Draw the entries of $G$ on and above the diagonal independently,
with
\[
    G_{ii}\sim\mathcal N(0,2),\qquad
    G_{ij}\sim\mathcal N(0,1)\quad(i<j),
    \qquad G_{ji}=G_{ij}\quad(i<j).
\]
We recall the Gaussian conditioning results in this normalization.

\begin{proposition}[Conditional Gaussian Queries]
\label[proposition]{prop:conditional-gaussian-queries}
Let $G$ be a real symmetric $d\times d$ matrix with independent centered
Gaussian entries on and above the diagonal, of variance two
on the diagonal and one above it. Consider an adaptive matrix--vector
procedure querying $G$, with internal randomness $R$ independent
of $G$. Let $\mathcal T$ consist of $R$ and the query vectors
and responses so far.
Let $Q$ be the orthogonal projector onto the span of these
query vectors.
\begin{enumerate}[(i)]
    \item \label{item:conditional-gaussian-queries:response}
    Conditional Response
    {\cite[Lemma~3.4 and Appendix~C.2]{SER18}}.
    If the next query $\bm u$ is a unit vector orthogonal to
    that span, then
    \[
        (I_d-Q)G\bm u\mid\mathcal T
        \sim\mathcal N(\bm0,I_d-Q+\bm u\bm u^\top).
    \]
    \item \label{item:conditional-gaussian-queries:block}
    Conditional Block {\cite[Lemma~C.1]{JPWZ21}}.
    The conditional mean is
    \[
        \E_G[G\mid\mathcal T]=GQ+QG-QGQ.
    \]
    Conditional on $\mathcal T$, the centered matrix has the law
    \[
        G-\E_G[G\mid\mathcal T]
        \stackrel d=(I_d-Q)G'(I_d-Q),
    \]
    where $G'$ is an independent copy of $G$.
\end{enumerate}
\end{proposition}

The responses determine $GQ$, since they determine the action
of $G$ on the query span. Symmetry then determines
$QG=(GQ)^\top$.

\begin{proposition}[Frobenius Inner Products Of Outer Products]
\label[proposition]{prop:frobenius-outer-products}
For real vectors $\bm u_i$ and real scalars $x_i$,
\[
    \left\|\sum_i x_i\bm u_i\bm u_i^\top\right\|_F^2
    =\sum_{i,j}x_ix_j(\bm u_i^\top\bm u_j)^2.
\]
\end{proposition}

\begin{proof}
For any two real vectors $\bm u,\bm v$,
\[
    \langle\bm u\bm u^\top,\bm v\bm v^\top\rangle_F
    =\sum_{i,j}u_iu_jv_iv_j
    =\left(\sum_i u_iv_i\right)^2
    =(\bm u^\top\bm v)^2.
\]
Expanding the squared Frobenius norm by bilinearity gives
the claimed identity.
\end{proof}

We are now ready to prove the matrix--vector query lower bound for simultaneously estimating the diagonal entries of a random matrix.
\begin{proof}[Proof of \Cref{lem:optimal-context-attribution:diagonal-lower-bound}]
We first determine the conditional distribution of the diagonal
given the query vectors and responses. We then show that the
probability of estimating every diagonal entry to error
$\sqrt{\log d}/(128d)$ is small.

Set $A=G/(64d)$. Consider a procedure using at most
$d/8$ real matrix--vector queries, and run it on $A$.
A query to $A$ reveals the corresponding matrix--vector
product with $G$ by scaling the response by $64d$.

\smallskip
\noindent\emph{Conditional Diagonal.}
Condition on $\mathcal T$. By
\Cref{item:conditional-gaussian-queries:block}, jointly for
all $i\in[d]$,
\[
    G_{ii}-\E_G[G_{ii}\mid\mathcal T]
    \stackrel d=
    ((I_d-Q)\bm e_i)^\top G'((I_d-Q)\bm e_i),
\]
with the same $G'$ for every $i$.
For real vectors $\bm u,\bm v$, independence of the entries
on and above the diagonal gives
\begin{align*}
    \Cov_{G'}(\bm u^\top G'\bm u,\bm v^\top G'\bm v)
    &=2\sum_i u_i^2v_i^2
      +4\sum_{i<j}u_iu_jv_iv_j\\
    &=2(\bm u^\top\bm v)^2.
\end{align*}
Here a diagonal entry has variance two, and each off-diagonal
entry appears twice in each quadratic form. Taking
$\bm u=(I_d-Q)\bm e_i$ and $\bm v=(I_d-Q)\bm e_j$, and using
$(I_d-Q)^2=I_d-Q$, gives
\[
    \Cov_G(G_{ii},G_{jj}\mid\mathcal T)
    =
    \begin{cases}
        2(1-Q_{ii})^2,&i=j,\\
        2Q_{ij}^2,&i\ne j.
    \end{cases}
\]

Let $\mathcal I=\{i:Q_{ii}\le \nicefrac14\}$.
Since $\sum_iQ_{ii}=\operatorname{rank}(Q)\le \nicefrac{d}8$, we have
\[
    |\mathcal I|\ge d-4\sum_iQ_{ii}\ge \frac{d}2.
\]
Write $\Sigma=\Cov_G((G_{ii})_{i\in\mathcal I}\mid\mathcal T)$.
Since $Q$ is an orthogonal projector,
$(Q\bm e_i)^\top(Q\bm e_j)=Q_{ij}$.
Applying \Cref{prop:frobenius-outer-products} with
$\bm u_i=Q\bm e_i$ gives
\begin{align*}
    \bm x^\top(\Sigma-I_{|\mathcal I|})\bm x
    &=\sum_{i\in\mathcal I}(1-4Q_{ii})x_i^2
      +2\sum_{i,j\in\mathcal I}x_ix_jQ_{ij}^2\\
    &=\sum_{i\in\mathcal I}(1-4Q_{ii})x_i^2
      +2\left\|\sum_{i\in\mathcal I}
        x_i(Q\bm e_i)(Q\bm e_i)^\top\right\|_F^2
      \ge0
\end{align*}
for every real vector $\bm x$ indexed by $\mathcal I$.
The last inequality uses $Q_{ii}\le1/4$ on $\mathcal I$.
Thus $\Sigma\succeq I_{|\mathcal I|}$.

\smallskip
\noindent\emph{Simultaneous Accuracy.}
The internal randomness $R$ and the responses in $\mathcal T$
fix each estimate $\widehat L_i$. Conditional on $\mathcal T$,
independently draw
\begin{align*}
    \bm b&\sim\mathcal N\left(
      \E_G[(G_{ii})_{i\in\mathcal I}\mid\mathcal T],
      \Sigma-I_{|\mathcal I|}\right),\\
    \bm Z&\sim\mathcal N(\bm0,I_{|\mathcal I|}).
\end{align*}
The sum $\bm b+\bm Z$ is Gaussian with the conditional mean
of the diagonal and covariance
$(\Sigma-I_{|\mathcal I|})+I_{|\mathcal I|}=\Sigma$.
Thus, in this representation,
$A_{ii}=(b_i+Z_i)/(64d)$ for $i\in\mathcal I$.

Fix $\mathcal T$ and $\bm b$. The estimate $\widehat L_i$ is
fixed, while $Z_i$ is standard normal. Therefore
\[
    |\widehat L_i-A_{ii}|\le\frac{\sqrt{\log d}}{128d}
    \quad\Longleftrightarrow\quad
    |Z_i-(64d\widehat L_i-b_i)|\le\frac{\sqrt{\log d}}2.
\]
The interval on the right has the largest Gaussian probability
when centered at zero, because the standard Gaussian density
is symmetric and decreases with distance from zero.
For a standard normal $Z$, independence of the $Z_i$ gives
\begin{align*}
    &\Pr_{\bm Z}\left(
      \forall i\in\mathcal I,\;
      |Z_i-(64d\widehat L_i-b_i)|\le\frac{\sqrt{\log d}}2
      \,\middle|\,\mathcal T,\bm b\right)\\
    &\qquad=
      \prod_{i\in\mathcal I}
      \Pr_Z\left(|Z-(64d\widehat L_i-b_i)|
          \le\frac{\sqrt{\log d}}2\right)\\
    &\qquad\le
      \Pr_Z\left(|Z|\le\frac{\sqrt{\log d}}2\right)^{|\mathcal I|}.
\end{align*}
The bound holds for every $\bm b$. Averaging over $\bm b$,
using $|\mathcal I|\ge d/2$ and the fact that success on all
coordinates requires success on $\mathcal I$, gives
\begin{align*}
    &\Pr_G\left(
      \max_{i\in[d]}|\widehat L_i-A_{ii}|
      \le\frac{\sqrt{\log d}}{128d}
      \,\middle|\,\mathcal T\right)\\
    &\qquad\le
      \left(1-\Pr_Z\left(|Z|>\frac{\sqrt{\log d}}2\right)\right)^{d/2}\\
    &\qquad\le
      \exp\left(-\frac{d}2
        \Pr_Z\left(|Z|>\frac{\sqrt{\log d}}2\right)\right)
      \alignnote{$1-x\le e^{-x}$}.
\end{align*}
Here $Z\sim\mathcal N(0,1)$.

It suffices to show that the exponent grows with $d$.
For $t\ge0$, integrating the standard Gaussian density over
$[t,t+1]$ gives
\begin{align*}
    \Pr_Z(|Z|>t)
    &\ge\frac1{\sqrt{2\pi}}\int_t^{t+1}e^{-z^2/2}\,dz\\
    &\ge\frac{e^{-(t+1)^2/2}}{\sqrt{2\pi}}
     \ge\frac{e^{-1-t^2}}{\sqrt{2\pi}}
     \alignnote{$(t+1)^2/2\le t^2+1$}.
\end{align*}
At $t=\sqrt{\log d}/2$, this gives
\[
    d\Pr_Z\left(|Z|>\frac{\sqrt{\log d}}2\right)
    \ge\frac{d^{3/4}}{e\sqrt{2\pi}}.
\]
It remains to find a matrix in the Frobenius ball on which the
procedure fails. For $d\ge64$, the preceding bound gives
\[
    \Pr_G\left(
      \max_{i\in[d]}|\widehat L_i-A_{ii}|
      \le\frac{\sqrt{\log d}}{128d}
      \,\middle|\,\mathcal T\right)
    \le\exp\left(-\frac{d^{3/4}}{2e\sqrt{2\pi}}\right)
    \le\exp\left(-\frac8{e\sqrt\pi}\right)<\frac14.
\]
Averaging over the transcript gives the same bound on the
unconditional success probability.

Finally, $\E_G\norm{G}_F^2=d(d+1)$.
By Markov's inequality,
\[
    \Pr_G\left(\norm{A}_F>\frac1{16}\right)
    =\Pr_G(\norm{G}_F>4d)
    \le\frac{d(d+1)}{16d^2}\le\frac14.
\]
Thus, with probability at least $1-1/4-1/4=1/2$, the matrix
lies in the ball and some diagonal estimate has error greater
than $\sqrt{\log d}/(128d)$. Consequently, some fixed matrix
in the ball has failure probability at least $1/2$ over the
procedure's internal randomness $R$.
\end{proof}

\end{document}

%% file: headers.tex
\usepackage{parskip}
\usepackage[margin=1in]{geometry}
\usepackage[utf8]{inputenc}
\usepackage[T1]{fontenc} %
\usepackage{mathpazo}
\usepackage{bm}
\usepackage{physics}
\usepackage{nicefrac}
\usepackage{mathtools}
\usepackage{alignnotes}

\usepackage{upgreek}
\renewcommand{\gamma}{\upgamma}
\renewcommand{\pi}{\uppi}
\newcommand{\eps}{\varepsilon}
\newcommand{\ii}{\mathrm{i}\mkern1mu}

\newcommand{\eg}{\emph{e.g.}}

\newcommand{\R}{\mathbb{R}}
\newcommand{\C}{\mathbb{C}}

\DeclareMathOperator*{\E}{\mathbb{E}} 
\DeclareMathOperator*{\Var}{Var}
\DeclareMathOperator*{\Cov}{Cov}

\DeclareMathOperator*{\poly}{poly}
\DeclareMathOperator{\Trim}{Trim}

\newcommand{\Ot}{\widetilde{O}}

\DeclarePairedDelimiter{\iprod}{\langle}{\rangle}

\usepackage{amsthm}
\usepackage[ruled,vlined,linesnumbered]{algorithm2e}

\usepackage{markdown}
\markdownSetup{
  rendererPrototypes = {
    headingOne    = {\section*{#1}},
    headingTwo    = {\subsection*{#1}},
    headingThree  = {\subsubsection*{#1}},
    headingFour   = {\paragraph{#1}},
    headingFive   = {\subparagraph{#1}},
    headingSix    = {\textbf{#1}\par},
  }
}

\usepackage{cancel}
\usepackage{mathtools}

\usepackage{manfnt} %

\usepackage{figchild}
\usepackage{fontawesome5}

\usepackage{accents}
 
\usepackage[dvipsnames]{xcolor}
\usepackage{color} 
\definecolor{niceRed}{RGB}{190,38,38}
\definecolor{niceYellow}{HTML}{f5b400}
\definecolor{blueGrotto}{HTML}{059DC0}
\definecolor{royalBlue}{HTML}{057DCD}
\definecolor{navyBlue}{HTML}{0B579C}
\definecolor{yaleBlue}{HTML}{00356b}
\definecolor{limeGreen}{HTML}{81B622}
\definecolor{nicePurple}{HTML}{9c27b0}
\definecolor{lightRoyalBlue}{HTML}{def2ff}  
\definecolor{gold}{HTML}{ffa300}
\definecolor{claudeBlue}{HTML}{4C72B0}
\definecolor{claudeRed}{HTML}{C44E52}

\usepackage[most]{tcolorbox}

\newtcolorbox{framed}[1][]{%
  breakable,
  arc=4pt, boxrule=0.6pt,
  colframe=black, colback=white,
  left=6pt, right=6pt, top=4pt, bottom=4pt,
  before skip=8pt, after skip=8pt,
  fonttitle=\bfseries, coltitle=black,
  attach title to upper={\ }, %
  title={#1},
}

\usepackage{xcolor}
\usepackage{xfp} %
\usepackage{graphicx}
\usepackage{subcaption}
\usepackage{hyperref}
\hypersetup{
  colorlinks = true, 
  urlcolor = {blueGrotto},
  linkcolor = {royalBlue},
  citecolor = {navyBlue}
}

\usepackage[capitalise,noabbrev,sort,nameinlink]{cleveref}

\newtheorem{theorem}{Theorem}[section]
\usepackage{thmtools}
\declaretheorem[sibling=theorem]{proposition}
\declaretheorem[sibling=theorem]{lemma}

\declaretheorem[sibling=theorem]{example}

\makeatletter
\let\statementparts@originalresetby\cref@resetby
\newcommand*\statementparts@resetby[2]{%
  \ifstrequal{#1}{enumi}
    {\def#2{theorem}}
    {\statementparts@originalresetby{#1}{#2}}%
}
\newcommand*\statementpartssetup[1]{%
  \crefalias{enumi}{#1}%
  \let\cref@resetby\statementparts@resetby
  \apptocmd{\@enumerate@}{%
    \ifnum\@enumdepth=1\relax
      \renewcommand{\p@enumi}{\thetheorem}%
      \renewcommand{\theenumi}{(\roman{enumi})}%
      \renewcommand{\labelenumi}{(\roman{enumi})}%
    \fi
  }{}{%
    \PackageError{statementparts}
      {Cannot configure statement parts}
      {Check the enumerate implementation.}%
  }%
}
\newcommand*\EnableStatementPartReferences{%
  \AtBeginEnvironment{theorem}{\statementpartssetup{theorem}}%
  \AtBeginEnvironment{proposition}{\statementpartssetup{proposition}}%
  \AtBeginEnvironment{lemma}{\statementpartssetup{lemma}}%
  \AtBeginEnvironment{corollary}{\statementpartssetup{corollary}}%
}
\makeatother

\usepackage[
  backref=true,
  backend=biber,
  natbib=true,
  style=alphabetic,
  sorting=alphabeticlabel,
  sortcites=true,
  minbibnames=3,
  maxbibnames=999,
  mincitenames=4,
  maxcitenames=4,
  minalphanames=4,
  maxalphanames=4,
  url=false,
  doi=false,
]{biblatex}

\DeclareSortingTemplate{alphabeticlabel}{
  \sort[final]{%
    \field{labelalpha} %
  }
  \sort{%
    \field{year}   %
  }
  \sort{%
    \field{title}  %
  }
}

\AtBeginRefsection{\GenRefcontextData{sorting=ynt}}
\AtEveryCite{\localrefcontext[sorting=ynt]}